\documentclass[prX,twocolumn,showpacs,preprintnumbers,amsmath,amssymb,longbibliography]{revtex4-1} 

\usepackage{dcolumn}
\usepackage{bm}

\usepackage{amsmath}
\usepackage{amssymb}
\usepackage{amsfonts} 
\usepackage{wasysym}  
\usepackage{graphics}  
\usepackage{psfrag} 
\usepackage{graphicx} 
\usepackage{epsfig}    
\usepackage{rotating}  
 \usepackage[latin1]{inputenc}
 \usepackage{color}
 \usepackage{rotating}
\usepackage{csquotes}
 \usepackage{multirow}
\usepackage[nooneline, tight,raggedright,FIGTOPCAP]{subfigure}
\usepackage[linkcolor=red,citecolor=blue,urlcolor=blue,colorlinks=true]{hyperref}
\usepackage{color}
\usepackage{epsfig}
\usepackage[normalem]{ulem}
\definecolor{darkgreen}{rgb}{0,0.5,0} 
\definecolor{violet}{rgb}{0.5,0,0.5}
\definecolor{orange}{rgb}{0.2,0.5,0.5}
\newcommand{\avg}[1]{\langle #1\rangle}

\newcommand{\bequ}{\begin{equation}}
\newcommand{\eequ}{\end{equation}}
\newcommand{\bequa}{\begin{eqnarray}}
\newcommand{\eequa}{\end{eqnarray}}
\newcommand{\bse}{\begin{subequations}}
\newcommand{\ese}{\end{subequations}}
\newcommand{\tn}[1]{\textnormal{#1}}
\newcommand\est{\bgroup \markoverwith{\textcolor{darkgreen}{\rule[0.5ex]{2pt}{1.5pt}}}\ULon}
\newcommand\eul{\bgroup \markoverwith{\textcolor{darkgreen}{\rule[-0.5ex]{2pt}{1.5pt}}}\ULon}

\begin{document}


\title{
Two-species active transport along cylindrical biofilaments is limited by emergent topological hindrance
}

\author{Patrick Wilke}
\thanks{P.W. and E.R. contributed equally to this work.}
\author{Emanuel Reithmann}
\thanks{P.W. and E.R. contributed equally to this work.}
\author{Erwin Frey}
\email{frey@lmu.de}

\affiliation{
Arnold Sommerfeld Center for Theoretical Physics (ASC) and Center for NanoScience (CeNS), Department of Physics, Ludwig-Maximilians-Universit\"at M\"unchen, Theresienstrasse 37, 80333 M\"unchen, Germany and Nanosystems Initiative Munich (NIM), Ludwig-Maximilians-Universit\"{a}t M\"{u}nchen, Schellingstra{\ss}e 4, 80333 Munich, Germany
}

\begin{abstract}
Active motion of molecules along filamentous structures is a crucial feature of cell biology and is often modeled with the paradigmatic asymmetric simple exclusion process. Motivated by recent experimental studies that have addressed the stepping behavior of kinesins on microtubules, we investigate a lattice gas model for simultaneous transport of two species of active particles on a cylinder. The species are distinguished by their different gaits: While the first species moves straight ahead, the second follows a helical path. We show that the collective properties of such systems critically differ from those of one-species transport in a way that cannot be accounted for by standard models. This is most evident in a jamming transition far below full occupation, as well as in non-equilibrium pattern formation. The altered behavior arises because -- unlike the case in single-species transport -- any given position may be targeted by two particles from different directions at the same time. However, a particle can leave a given position only in one direction. This simple change in connectivity significantly amplifies the impact of steric interactions and thus becomes a key determinant of mixed species transport. We computationally characterize this type of hindrance and develop a comprehensive theory for collective two-species transport along a cylinder. Our observations show high robustness against model extensions that account for additional biomolecular features and demonstrate that even small fractions of a second species can significantly alter transport. This suggests that our analysis is also relevant in a biological context.
\end{abstract}

\date{\today}

\maketitle

\section{Introduction and motivation}
Efficient collective molecular transport is a vital prerequisite for a multitude of processes in cell biology on many different levels. Examples range from mRNA translation to organelle transport. Typically, highly functional molecular motors which transform chemical energy into stepwise mechanical translation perform this complex task by moving along filamentous structures such as the cytoskeleton or mRNA~\cite{Howard2001,Nilsson2005}.  Of particular importance for intracellular organization are the cylindrically shaped and polarized microtubules. Kinesins -- the molecular motors associated with microtubules --  exhibit distinct efficient motility, as they are capable of ``walking" processively over micrometer distances towards the microtubule end~\cite{Howard2001}. 
Many experimental studies have focused on elucidating the microscopic working mechanisms of  molecular motors~\cite{Woehlke2000,Cross2004,Gennerich2009}, but understanding their collective behavior~\cite{Leduc2012} remains a challenging task. For this reason,  concepts drawn from statistical physics and modeling have proven to be of much relevance, as they offer a means of linking the microscopic to macroscopic behavior. In this context, the totally asymmetric exclusion process (TASEP)~\cite{MacDonald1968,Chou2011b} and extensions thereof have proven particularly fruitful. In its simplest formulation, the TASEP accounts for two central aspects of transport: active motion and steric interactions. It describes the motion of point particles along a one-dimensional lattice, and therefore serves as an ideal basis to study collective transport of one particle species along a single track. Despite its simplicity, this model shows a surprisingly rich phenomenology and captures many essential features of transport processes. Indeed, it has by now acquired the status of a paradigmatic model, not only for transport~\cite{Chou2011b,Mobilia2006} but also for non-equilibrium physics in general, comparable to that of the Ising model  for equilibrium physics.

Motion of kinesins was initially thought to occur mostly along the so-called protofilaments --  separate lanes oriented parallel to the axis of the cylindrical microtubule  (see Fig.~\ref{fig:model}, green kinesin).  While this motion along a single protofilament is a well studied feature of one of the most prominent kinesins involved in intracellular transport, kinesin-1~\cite{Ray1993}, several \emph{in vitro} studies have revealed notable exceptions to this behavior. Members of all three superfamilies of molecular motors, kinesin~\cite{Yajima2005,Walker1990,Pan2010,Yajima2008}, dynein~\cite{Vale1988}, and myosin~\cite{Nishizaka1993}, have been shown  to produce torsional force during their axial translation. While the precise basis for this observation has remained unclear, recent studies of kinesin-2~\cite{Brunnbauer2012} and the microtubule depolymerizing kinesin-8~\cite{Bormuth2012,Bugiel2015} strongly suggest that these molecular motors regularly switch protofilaments and show a bias towards one side. This effectively results in a helical motion along the microtubule~\cite{Brunnbauer2012,Can2014,Stepp2017} (see Fig.~\ref{fig:model}, orange kinesin). On a theoretical level, systems with a single species of kinesins that stochastically switch protofilaments have been studied recently by employing extended TASEP-like models with multiple lanes~\cite{Curatolo2015a}. In that case, the qualitative behavior of collective transport along the cylindrical microtubule was reported to be widely conserved as compared to the case of protofilament tracking kinesins. However, many different types of molecular motors are present in a single cell and, as lane switching is likely to be the rule rather than the exception, the general scenario is that several  different molecular motor species should interact on a single cytoskeletal filament. This raises the question of  how the interplay of distinct molecular motor species that  show different gaits alters collective transport along a cylindrical structure.

Inspired by molecular transport on microtubules we employ lattice gas models to find generic principles of collective transport by two species of particles that are distinguished by different gaits  on a cylinder.
Here, we demonstrate that the emerging behavior of such systems critically differs from collective behavior in the presence of a single species only. The simultaneous presence of molecular motors that follow a straight and a helical course, respectively, inevitably leads to crossings between their trajectories: A certain lattice site may be targeted from two different directions, but once occupied can only be vacated in a single direction. This modification of connectivity in the network topology of potential particle movements amplifies the impact of steric interactions globally, and thus hinders particle motion -- an effect we call \emph{topological hindrance}. Topological hindrance produces highly non-trivial correlations between the dynamics of particles of the different species. Specifically, the particle current and distribution on the filament are now dependent not only on the total number of particles, but also on the fraction of the respective species. The impact of topological hindrance is most evident in the jamming of particle flows at densities far below full occupation. We present an analytical framework that quantifies topological hindrance and provides a theoretical basis for understanding two-species transport along cylinders, much as the TASEP does for transport by a single motor species. Moreover, our model predicts non-equilibrium patterns in the particle distribution that have not been observed in classical models for single-species transport. To specifically target the robustness and  biological relevance of topological hindrance we further investigate extended models that account for specific biomolecular features. We find that topological hindrance still plays a key role for collective transport properties in these cases. While the extended models are  too complicated for an analytic investigation we can understand their behavior on the basis of our idealized model and our theory for topological hindrance.

This manuscript is organized as follows: We begin with a review of the collective transport by a single species in Section~\ref{sec:TASEP}. Our model for collective transport by two species on a cylindrical structure is presented in Section~\ref{sec:model}. The phenomenology of our model and the key differences to transport with a single species are discussed in Section~\ref{sec:tophind}, which also provides a qualitative explanation of topological hindrance. Further we address how the system dimensions influence topological hindrance. We then develop a theory to quantify the strength of topological hindrance for arbitrary particle densities in Section~\ref{sec:current_density} which yields the current-density relation for an arbitrary number of lanes  and species fractions. This allows us to compute the complete phase diagram of our model, i.e. the particle density and current emerging in the system as a function of the control parameters. In Section~\ref{sec::biological_relevance} we discuss the robustness of our results against model modifications and their biological relevance, and provide a guideline for potential experimental verification. Finally, in Section~\ref{sec:conclusion}, we relate our work to existing mathematical theories of driven systems, and discuss its applicability and possible relevance in the biological context.

\section{Revisiting one-species transport}\label{sec:TASEP}
We start with a summary of the TASEP, which is one of the most fundamental models used to describe active transport of sterically interacting agents along defined pathways. While exact results relating to its properties have had a major impact on the field of non-equilibrium physics in general~\cite{Derrida1992,Derrida1993a, Schutz1993, Schutz2001, Gorissen2012}, its applications cover a vast variety of transport processes such as ion channels~\cite{Katz1983,Chou1998}, spin transport~\cite{Reichenbach2006,Reichenbach2007}, traffic flow~\cite{Schadschneider2011}, mRNA translation~\cite{Chou2004}, and intracellular transport~\cite{Chou2011b}. Here, the TASEP will serve as the starting point to treat molecular transport along cytoskeletal structures in the presence of a single molecular motor species. 

The model is defined as follows: Point particles populate a one-dimensional lattice along which they can hop stochastically at a rate $\nu$ to the neighboring lattice site on their right. To account for steric interactions, the particles exclude each other, such that a lattice site can only be occupied by a single particle. The particles enter the lattice from the left at a rate $\alpha$ and leave the lattice on the right at a rate $\beta$. This can be interpreted as connecting the lattice with two particle reservoirs with fixed densities $\rho_\mathrm{L}=\alpha$ on the left and $\rho_\mathrm{R}=1-\beta$ on the right.

With these definitions, the TASEP allows one to study macroscopic properties that emerge in transport processes. Two central observables are the average particle density $\rho(x,t)$ (average particle distribution) and the average particle flux $J(x,t)$, where $x$ denotes the position in the system and $t$  the time. Particles can not be created or annihilated within the lattice. Therefore, a spatial difference in the particle flux must lead to a temporal change in the particle distribution in the ensemble average.  This is reflected by a continuity equation that describes the system's temporal evolution on a macroscopic scale~\footnote{Note that in the macroscopic limit $x$ becomes a continuous variable and the lattice is treated as a continuum.}, 
\bequa
	\partial_t \rho(x,t) + \partial_x J(x,t)= 0.
\eequa
Both the TASEP and the model considered in this manuscript are ergodic and the Perron-Fobenius theorem holds true~\cite{Seneta2006}. This means that they will evolve into a unique non-equilibrium steady state, on which we will focus from now on.

For the TASEP and many other transport models, there is a unique connection between the current and the density, the current-density relation $J(\rho)$.  The existence of this unique function means that the local current $J$ is completely determined by the density $\rho$; $J$ depends on the in rate $\alpha$ and out rate $\beta$ only implicitly via the density. This phenomenological approach based on a unique current-density relation goes back to the work of Lighthill and Whitham~\cite{Lighthill1955} and is at the heart of many theories for various transport models~\cite{Krug1991,Popkow1999,Blythe2007, Curatolo2015a}. For the TASEP, the current-density relation is given by~\cite{Derrida1992,Derrida1993a}
\bequa
\label{eq:current_density_TASEP}
	J(\rho) = \nu \rho (1- \rho). 
\eequa
This equation reflects the fact that particles may only move to empty lattice sites: Neglecting correlations, the probability of finding a particle (given by $\rho$) must be multiplied by the probability that a lattice site is empty (given by $1-\rho$) to obtain the current. Although this is just a heuristic mode of argumentation, it can be proven that the corresponding Eq.~\ref{eq:current_density_TASEP} is an exact relation for the TASEP on an infinite dimensional lattice~\cite{Derrida1992,Derrida1993a, Schutz1993}. This relation also implies that the particle current vanishes if the density is either zero (no particles) or one (full occupation). In the latter case, the motion of every particle except for the last one is blocked, and  consequently no hopping within the lattice can occur. We refer to such a system as \emph{jammed}. 

Based on the current-density relation, it is possible to describe the complete macroscopic behavior in terms of the control parameters of the TASEP in the stationary state. Specifically, it uniquely determines the average density and particle current of the stationary state associated with a specific choice of control parameters $\alpha$ and $\beta$.
A very successful theoretical concept in this context is the extremal current principle~\cite{Krug1991,Popkow1999,Hager2001}. It states that, given a set of possible densities, the system will always realize the one corresponding to the extremal current. As explained above, the lattice boundaries are effectively connected to particle reservoirs of density $\rho_\mathrm{L}=\alpha$ and $\rho_\mathrm{R}=1-\beta$ on the left and right end, respectively. According to the extremal current principle, the steady state density and current are then given by
\bequa
	\label{eq:ECP}
	J=\begin{cases}
    \max\limits_{\rho  \in [\rho_R,\rho_L]} J(\rho) & \tn{if $\rho_L > \rho_R$}\\
    \min\limits_{\rho  \in [\rho_L,\rho_R]} J(\rho) & \tn{if $\rho_L < \rho_R$} .
  \end{cases}
\eequa
The current-density relation of the TASEP, Eq.~\ref{eq:current_density_TASEP}, exhibits only a single maximum. Consequently, the extremal current principle predicts boundary induced transitions between three different phases: Eq.~\ref{eq:ECP} implies that either the left reservoir density $\rho_L$, the right reservoir density $\rho_R$, or the density corresponding to the sole maximum in the current-density relation,  $\rho_\mathrm{MC}=0.5$, are valid solutions and therefore realized in the system. On an intuitive level, when the left reservoir density is realized, a lack of particles determines the system's behavior, as the in rate is too small to create large particle jams. The corresponding phase is called low-density (LD) phase. When the out rate is so small as to be the factor that limits transport, particles start to jam at the lattice end. Ultimately, the corresponding outflux will determine the overall particle current. The system then realizes the right reservoir density. This  phase is called high-density (HD) phase. Third, if neither a lack of particles nor the out rate are limiting transport, the particles' motion itself and thus the maximal possible current and the corresponding density constitute a constraint. This phase is the maximal-current (MC) phase.

It is worth noting that the above derivation of phase transitions is independent of the microscopic rules of a system and has also been applied successfully to driven lattice gas models other than the TASEP~\cite{Hager2001,Melbinger2012}. Therefore, the extremal current principle suggests that the phase behavior of a driven lattice gas is qualitatively identical to that of the TASEP as long as the curve characteristics of the current-density relation are conserved: A region where the current shows a monotonic increase with increasing density, a region where the current monotonically decreases with increasing density and a single maximum. 

The TASEP has been generalized in various ways. Several studies investigated systems with a more involved geometry such as two lanes \cite{Popkov2003a,Pronina2004,Reichenbach2006,Reichenbach2007,Evans2011},  junctions~\cite{Brankov2004,Pronina2005,Wang2008,Liu2009,Embley2009,Hilhorst2012,Ding2011}  or networks~\cite{Klumpp2005,Embley2008,Neri2011,Ezaki2012b,Neri2013a}. In particular, Curatolo et al.~\cite{Curatolo2015a} treated TASEPs with an arbitrary number of parallel lanes on a cylinder, where particles are also allowed to switch lanes. The authors find that in this case the system reduces in many ways to the single lane TASEP, which justifies its application for the case of a single species of lane-switching molecular motors on a microtubule. Besides these studies addressing more complicated geometries, several authors have treated multiple species of particles, typically in opposite directions~\cite{Evans1995,Schutz2003,Klumpp2004a,Blythe2007,Juhasz2007,Prolhac2009,Johann2014}. Yet, how a mixture of different species of molecular motors--that naturally may move with different gaits--behave collectively on the cylindrical microtubule structure remains elusive.

\section{Modeling two-species transport\label{sec:model}}
\begin{figure}[!b]
\centering
\underline{}\includegraphics[width = \columnwidth]{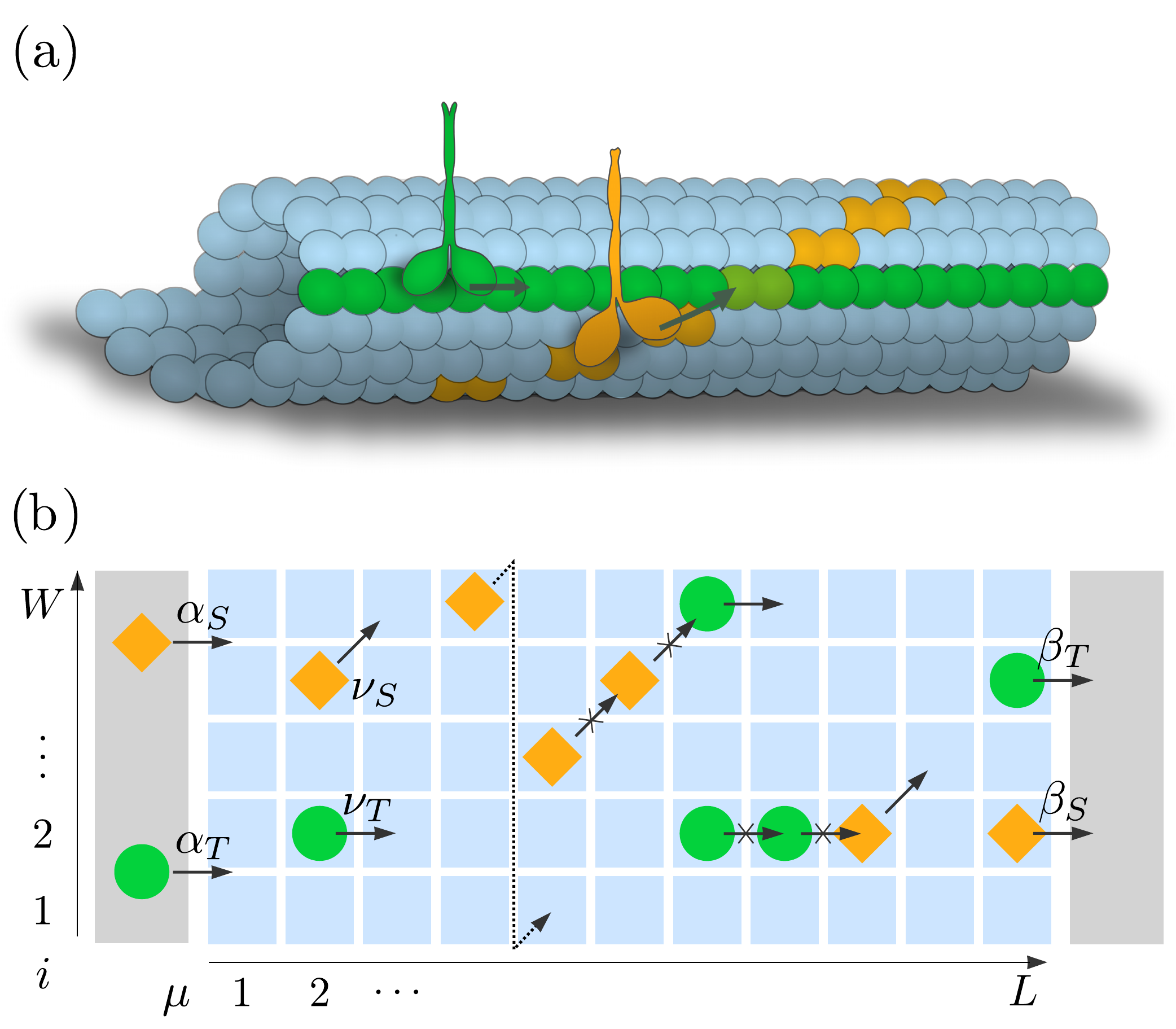}
\caption{\textbf{Active transport by two species along a cylindrical structure.} \textbf{(a)} Kinesins (e.g.\ kinesin-1, green) may track a single lane of a microtubule (protofilament) but many molecular motor species have also been reported to regularly switch protofilaments in a biased fashion and thereby effectively undergo a helical motion (e.g.\ kinesin-2, orange).
\textbf{(b)} Model implementation to study collective two-species transport along cylindrical structures. We consider a lattice with length $L$ and width $W$. Periodic boundary conditions  are employed along the transversal direction to account for a cylindrical geometry (e.g. microtubule). Particles of species $T$ (lane tracking, green) hop to right neighboring lattice sites at rate $\nu_T$, and particles of species $S$ (lane switching, orange) hop to the upper-right neighboring lattice site at rate $\nu_S$. All particles exclude each other. Particles enter the system with rates $\alpha_T$ and $\alpha_S$ at the left boundary and exit the system at the right boundary with rates $\beta_T$ and $\beta_S$, respectively. Gray areas denote system boundaries.}

\label{fig:model}
\end{figure}
As discussed in the previous section, the TASEP adequately describes molecular transport along a cylinder in the presence of a single species of molecular motors. We now turn to a mixture of two species of molecular motors that are distinguished by different gaits. Specifically, we will address the question of how to describe collective transport in the presence of molecular motors that move parallel to the cylinder axis \emph{and} molecular motors that follow a helical path as suggested by experiments~\cite{Brunnbauer2012,Bormuth2012,Bugiel2015}. To this end, we study the stochastic model with Markovian dynamics  illustrated in Fig.~\ref{fig:model}. We consider a two-dimensional lattice composed of  $W$ parallel lanes, each with a length of $L$ lattice sites. The system is populated by two different particle species: A \emph{lane-tracking} species ($T$ species) and a \emph{lane-switching} species ($S$ species).  Particles of the lane-tracking species stochastically hop at rate $\nu_T$ to the right neighboring lattice site while staying on the same lane. In detail, using Latin letters to denote the lane index and Greek letters to denote the site index, hopping of $T$ particles is described by $ i \rightarrow i$ and $\mu\rightarrow \mu+1$. They represent, for example, molecular motors that track a single protofilament. Particles of the $S$ species change lanes with every hopping event and stochastically move at rate $\nu_S$ to the upper-right neighboring lattice site, i.e. $i \rightarrow i+1$, and $\mu \rightarrow \mu+1$. To implement a cylindrical structure, periodic boundary conditions in the  transversal direction are imposed. Thus, members of the $S$ species hop from the uppermost to the lowermost lane.  In this way, the lane-switching particles represent molecular motors that move in spirals around the microtubule. 
Furthermore,  particles are subject to steric interactions. In the model, they exclude each other, and hopping events can only occur if the corresponding lattice site is empty. 
At the left boundary an empty site is filled with a particle of the respective species at rate $\alpha_T$ and $\alpha_S$. Conversely, at the right boundary particles of the $T$ and $S$ species leave the lattice at rates  $\beta_T$ and $\beta_S$.

Note that the assumptions of lane switching in each step and the absence of random particle attachment and detachment (Langmuir kinetics) are simplifications from a biological point of view.  Their aim is to isolate the basic principles of  two-species transport along a cylinder which, in turn, allows us to develop an analytic theory for topological hindrance. To bridge back to biological systems and prove relevance of our concepts, we discuss extended models in Section~\ref{sec::biological_relevance}.

To describe the state of the system, we use occupation numbers $n^T_{i,\mu},n^S_{i,\mu}\in \lbrace 0,1\rbrace $ for lattice site $\mu$ on lane $i$. Here  1 indicates that the  lattice site is occupied by a $T$ or $S$ particle,  whereas 0 stands for the absence of the respective species. We focus our analysis on the average particle distribution (density) $\rho$ and the average particle current $J$ that emerge in the system with the above stochastic rules. More specifically, we define $\rho^X_{i,\mu} := \avg{n^X_{i,\mu}}$ as the ensemble averaged occupation of site $\mu$ on lane $i$ by a particle of type $X\in \{T,S\}$. The current of $S$ particles at site $\mu$ on lane $i$ is defined as the average number of $S$ particles hopping onto this site per unit time: $J^S_{i,\mu}:= \nu_S \avg{n^S_{i-1,\mu-1} (1 - n^T_{i,\mu} - n^S_{i,\mu})}$. Equivalently, the current for the $T$ species $J^T$ is defined as $J^T_{i,\mu}:= \nu_T \avg{n^T_{i,\mu-1} (1 - n^T_{i,\mu} - n^S_{i,\mu})}$. Due to periodic boundary conditions along the transversal direction, we implicitly use the identification $i =W+1 \equiv 1$ and $i = 0 \equiv W$ in these relations and in the following. For later convenience we also define the occupation number irrespective of the particle species, $n_{i,\mu}:= n_{i,\mu}^T + n_{i,\mu}^S$. The temporal evolution of average occupations in the bulk of the system ($\mu \neq {1,L}$) is then given by the master equations
\bse
\label{eq:master_eq}
\bequa
 \frac{\mathrm{d}}{\mathrm{dt}} \avg{n^T_{i,\mu}} &=& \nu_T \avg{n^T_{i,\mu-1} (1 - n_{i,\mu} )} 
 - \nu_T \avg{n^T_{i,\mu} (1 - n_{i,\mu+1} )}  \nonumber \\
 &=& J^T_{i,\mu}-J^T_{i,\mu+1} , \label{eq:master_eqa} \\
 \frac{\mathrm{d}}{\mathrm{dt}} \avg{n^S_{i,\mu}} &=& \nu_S \avg{n^S_{i-1,\mu-1} (1 - n_{i,\mu})}   - \nu_S  \avg{n^S_{i,\mu} (1 - n_{i+1,\mu+1} )} \nonumber \\
  &=& J^S_{i,\mu}-J^S_{i+1,\mu+1}. \label{eq:master_eqb}
\eequa
\ese
At the boundary sites  $\mu=1,L$ the equations read
\bse
\label{eq:master_eq_bs}
\bequa
\frac{\mathrm{d}}{\mathrm{dt}} \avg{n^T_{i,1}} &=& \alpha_T (1-\avg{n_{i,1}})- J^T_{i,2}, \label{eq:master_in_T} \\
 \frac{\mathrm{d}}{\mathrm{dt}} \avg{n^S_{i,1}} &=& \alpha_S  (1-\avg{n_{i,1}})- J^S_{i+1,2} , \\
 \frac{\mathrm{d}}{\mathrm{dt}} \avg{n^T_{i,L}} &=& J^T_{i,L} - \beta_T \avg{n^T_{i,L}} , \\
 \frac{\mathrm{d}}{\mathrm{dt}} \avg{n^S_{i,L}} &=& J^S_{i,L} - \beta_S \avg{n^S_{i,L}}  . 
\eequa
\ese
By rescaling time,  it is possible to set  one hopping rate equal to $\nu_X = 1$ without loss of generality.  For simplicity we will focus on identical hopping rates $\nu_S=\nu_T=1$ throughout this manuscript. 

\section{The phenomenology of topological hindrance\label{sec:tophind}}

\subsection{Two-species transport can not be reduced to one-dimensional or single-species transport}
In this work, we focus on steady-state properties, i.e. $\mathrm{d} \avg{n^X_{i,\mu}}/\mathrm{dt}  =0$ and likewise for other moments. Eqs.~\ref{eq:master_eq} show that the average occupation of a lattice site depends on higher moments, which ultimately leads to an unclosed hierarchy of equations. This typically precludes an exact solution.
The simplest analytic approach to treat Eqs.~\ref{eq:master_eq} and \ref{eq:master_eq_bs}  is the mean-field approximation~\cite{Chou2011b}, where correlations are neglected by factorizing (second) moments, $\avg{n^X_{i,\mu} n^Y_{j,\nu}} = \avg{n^X_{i,\mu}} \avg{n^Y_{j,\nu}}$, which closes the hierarchy. The mean-field approximation works successfully for the TASEP and various similar models, and it therefore has acquired the status of a standard method for the analytical treatment of driven lattice gases~\cite{Chou2011b}. We employ this factorization scheme  for the total current $J_{i,\mu} := J^S_{i,\mu}+J^T_{i,\mu}$.  Furthermore, our system is irreducible  \footnote{It is possible to get from any arbitrary state to another arbitrary state in a finite number of steps by letting the system run empty and filling it with the respective configuration.}. For a continuous time Markov process this suffices to show that the stationary state is unique~\cite{Liggett2010}. Since there is only one stationary state it has to adapt the symmetry of the system. Thus, all macroscopic quantities have to be independent of the lane number  and we therefore omit the lane index $i$ in the following. This reasoning is further validated by stochastic simulations as shown in Appendix~\ref{ap:Rotational_Invariance}.  The mean-field current-density relation then reads
\bequa
\label{eq:current_density_mf}
 J_{\mu} &=& \rho^T_{\mu-1} (1-\rho^T_{\mu}-\rho^S_{\mu}) + \rho^S_{\mu-1} (1-\rho^T_{\mu}-\rho^S_{\mu}) \nonumber \\
  &=& \rho_{\mu-1} (1-\rho_{\mu}).
\eequa
Here we also defined the total particle density $\rho_\mu := \rho^T_\mu+\rho^S_\mu$ at site $\mu$. Eq.~\ref{eq:current_density_mf} is identical to the current-density relation of the TASEP, Eq.~\ref{eq:current_density_TASEP}. In particular, it predicts that the current is independent of the fraction of spiraling molecular motors and depends only on the total density. To relate the current and density to the system's control parameters it is useful to introduce new quantities. First, the total in rate given by $\alpha := \alpha_T + \alpha_S$. Second, since particles enter the system independently, the fraction of the current contributed by lane-switching particles is $\delta := \alpha_S/(\alpha_T+\alpha_S)$.  Due to current conservation  this current fraction is spatially constant.  In the mean-field analysis $J_S/J = \rho_S / \rho$ holds true, such that the current fraction $\delta$  of the $S$ species also equals the density fraction of $S$ particles. Therefore, we can consider $\rho$, $J$ instead of $\rho_S$, $\rho_T$, $J_S$, and $J_T$, and use $\delta$ to compute the respective fraction in the mean-field analysis.
Third, we can identify an effective out rate $\beta$ of particles irrespective of their species. The average dwell-time of a particle at the last site is $T= \delta /\beta_S + (1-\delta) /\beta_T$, and therefore $\beta := T^{-1} = (\beta_T \beta_S)/\left[ \delta \beta_T + (1-\delta ) \beta_S\right]$. With these definitions of $\alpha$ and $\beta$ we then obtain the full phase-diagram predicted by a na\"ive mean-field approximation, which recovers all TASEP relations. 

\begin{figure}[t]
\centering
\includegraphics[width = \columnwidth]{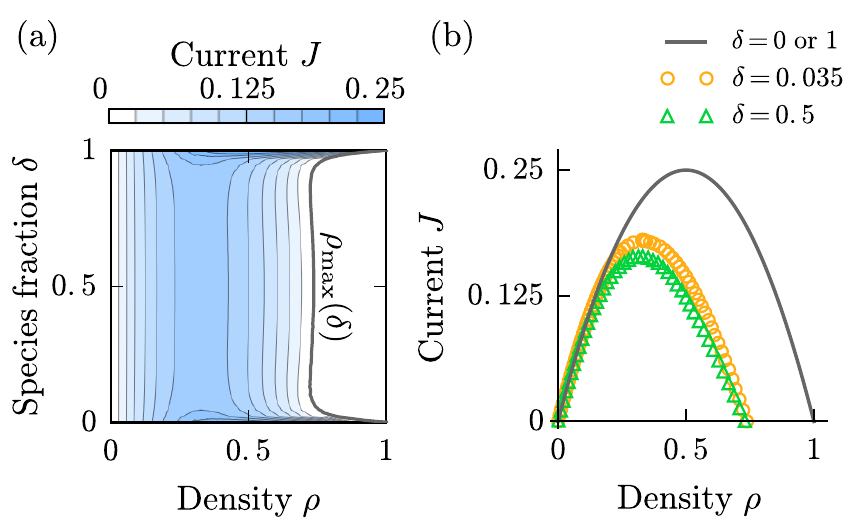}
\caption{\textbf{Current-density relation  obtained from stochastic simulations with $W=2$ lanes and $L=4096$ sites.}
 \textbf{(a)} The emerging current $J$ of an arbitrary lane (blue color) shows a dependence not only on the average system density $\rho$ but also on the fraction of lane-switching particles, $\delta$. The system jams at a maximal density $\rho_\mathrm{max}(\delta)$ (bold gray line); Densities between $\rho_\mathrm{max}$ and full occupation are not realized in any stationary state. \textbf{(b)} Current-density relation for mixed species ($\delta=0.5$, $\delta=0.035$, symbols) and single species ($\delta = 0$ or 1, line). For the latter cases, we recover the TASEP current-density relation. }
\label{fig:current_denisty}
\end{figure}
To test this mean-field analysis, we performed stochastic simulations based on Gillespie's algorithm~\cite{Gillespie2007} for a system with two lanes. Irrespective of the initial conditions the dynamics converge to a unique stationary state that is characterized by a particle density $\rho$ averaged over the whole system and a particle flux $J$ which we numerically computed for various values of $\alpha$, $\beta$, and $\delta$. The result for a system composed of two lanes is shown in Fig.~\ref{fig:current_denisty} (a) and (b). In clear contradiction to the mean-field analysis, Eq.~\ref{eq:current_density_mf}, we observe a strong dependence of the average current and density of particles on the fraction of spiraling molecular motors in the system. These findings falsify the mean-field approximation and show that the current is not uniquely determined by the density but carries an explicit dependence on the species fraction $\delta$. For fixed $\delta$, however, a  unique current-density relation $J(\rho(\alpha,\beta, \delta),\delta)$ can still be found, as shown in Fig.~\ref{fig:current_denisty} (a).  Unlike the species fraction $\delta$, the in rate $\alpha$ and out rate $\beta$ of particles are parameters that change the density only locally, and therefore the current only implicitly. 
The resulting current-density relation $J(\rho(\alpha,\beta, \delta),\delta)$, as shown for $W=2$ lanes in Fig.~\ref{fig:current_denisty} (a), exhibits a symmetry upon exchanging the species. This is explained by an invariance of particle dynamics when lattice sites are relabeled~\footnote{We can renumber the lane indices $j$ at position $i$ as $j\rightarrow j - (i \mod W)$.  This leads to a system with dynamical rules as the original system where lane switching particles now track lanes and vice-versa.}. For $\delta=0$ and $\delta=1$, i.e. in the presence of a single species only, we recover the TASEP current-density relation. If both particle species are present, the relation changes and the current is always less than in a single-species setup. Most interestingly, in a mixed system the current vanishes already at densities below full occupation (Fig.~\ref{fig:current_denisty} (a), white area); densities above a maximal value of $\rho_\mathrm{max}(\delta)$ (bold gray line) are not realized in the steady state. Moreover, already a very small fraction of lane-switching particles causes this effect: Approximately $2-5\%$ of a second species (see Fig.~\ref{fig:current_denisty}) are sufficient to cause significant deviations from the single-species and mean-field results. Besides this critical difference in the maximum density of a mixed and a one-species system, the qualitative shape of the current-density relation remains unchanged when the species fraction is varied. As shown in Fig.~\ref{fig:current_denisty} for the special case of $W=2$ lanes, we observe the same curve characteristics as for the TASEP. These are a single maximum for the current (MC), a low-density (LD) region where the current monotonically increases with increasing density and a high-density (HD) region where the current monotonically decreases with increasing density. The extremal current principle, Eq.~\ref{eq:ECP}, then suggests that  the phase diagram of our model and the TASEP are topologically equivalent although the positions of the phase boundaries might differ.

As the mean-field approximation fails to describe two-species transport, the species current fraction $\delta$ cannot be equated with the species density fraction. This complicates any further discussion as $\rho_S$ and $\rho_T$ have to be treated separately. However, as shown in Appendix~\ref{ap:species_fraction}, in the context of this manuscript no significant deviations of the density species fraction from the current species fraction occur. Therefore, for the remainder of this work, we mostly use $\rho$ and implicitly assume a conversion to $\rho_S$ and $\rho_T$ via $\delta$. Further, we refer to $\delta$ simply as species fraction.

In summary, our simulation data shows that, first, the dynamics of the two species are correlated, as manifested by the failure of the mean-field approach and the dependence of the current and density on the species fraction $\delta$. Second, correlations of the species dynamics always lead to a reduction of the average particle current; hence the presence of a second species always hinders transport. We conclude that our model can not be reduced to the TASEP and that na\"ive mean-field approaches are incapable of capturing the emerging macroscopic transport behavior.  

\subsection{Topological hindrance}
\begin{figure}[t]
\centering
\includegraphics[width = \columnwidth]{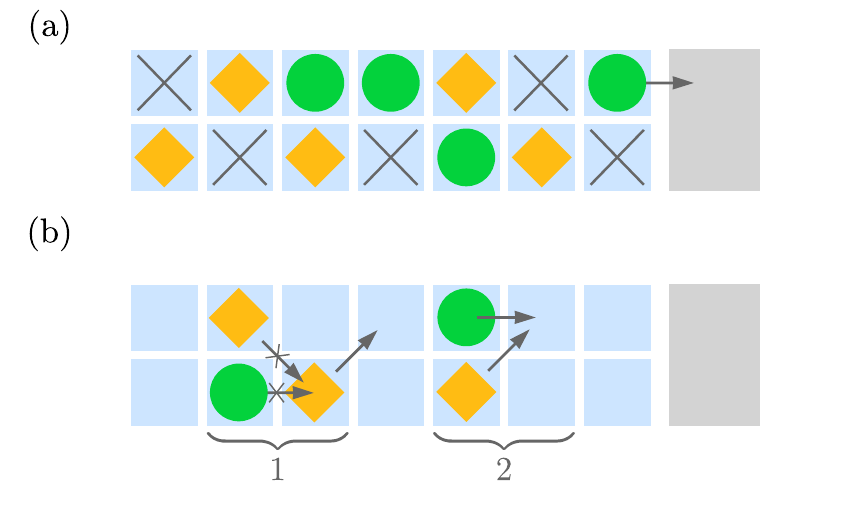}
\caption{\textbf{Illustration of topological hindrance for a system with $W=2$ lanes at the right lattice end.} \textbf{(a)} Shows a state where the motion of each particle in the bulk is blocked (jammed state) and only the rightmost particle is allowed to move (arrow). Empty lattice sites (crosses) are present, which lead to average densities below full occupation  for jammed systems. \textbf{(b)} The second species changes the connectivity of the network of possible particle motions which amplifies the impact of steric hindrance in our model. \emph{(1)} A single particle can now block up to two other particles.  \emph{(2)} Particles might align such that motion has to occur sequentially. Same symbols as in Fig.~\ref{fig:model} are used to illustrate the system and particles. }
\label{fig:hindrance_illustration}
\end{figure}

To obtain insight into the interactions that reduce the particle flux in the presence of both particle species, we take a qualitative look at particle configurations that might arise in our model. First, we consider jammed configurations where all particles (except for the very last one) are blocked and where, consequently, the average particle current vanishes. As discussed in Section~\ref{sec:TASEP}, for one species transport a jammed system can only arise trivially if every lattice site is occupied. In contrast our model exhibits jamming at densities below full occupation.
A possible jammed configuration of a mixed system is shown in Fig.~\ref{fig:hindrance_illustration} (a). 
Although the system is jammed, there exist lattice sites that are not occupied. These sites are inaccessible to any particle of this configuration (Fig.~\ref{fig:hindrance_illustration} (a), crosses) and ultimately lead to jamming below full occupation. 
In our model, two particles from different species can simultaneously target a single site. Consequently, if that site is occupied, a single particle can block the motion of two other particles (Fig.~\ref{fig:hindrance_illustration} (b), configuration 1). This, in turn, can create sites that are empty but nevertheless inaccessible. Moreover, if particles of both species try to access a single site that is empty and accessible (Fig.~\ref{fig:hindrance_illustration} (b), configuration 2) their motion is restricted in so far as either one of them but not both can hop at a time. These configurations hence act as an intrinsic bottleneck and will further reduce the particle current on average. 
To assess these considerations also qualitatively, we specifically computed the number of inaccessible lattice sites, and, with it, the maximal density $\rho_\mathrm{max}$ as a proof of principle for a system with $\delta=1/2$ and $W=2$ lanes. The detailed computations are shown in  Appendix~\ref{ap:rho_max} and lead to a value of $\rho_\mathrm{max}^\mathrm{theory} = 0.74$. This is in very good agreement with simulations that yield  $\rho_\mathrm{max}^\mathrm{sim} = 0.73$, which validates the qualitative arguments given above.

In summary, when the lane-switching species is added to the system, the network topology of possible particle motions is changed. Any given site is now potentially connected to two other sites although leaving a site is only possible in one direction at a time. This creates a two-to-one connectivity.  The single-species model, on the other hand, shows one-to-one connectivity or, for the general case of particles that may stochastically switch lanes, n-to-n connectivity.  The two-to-one connectivity of our model amplifies the impact of steric interactions relative to single-species transport and therefore hinders motion. First, steric interactions may now occur at more points in space as compared to single-species transport and thus give rise to inaccessible lattice sites (see Fig.~\ref{fig:hindrance_illustration} (b), group 1). Second, at intrinsic bottlenecks steric interactions can now act at more points in time, as particles may have to hop one after the other (see Fig.~\ref{fig:hindrance_illustration} (b), group 2). Throughout this manuscript, we refer to these phenomena as topological hindrance. 

\subsection{The influence of the number of lanes \label{sec:lane_dependence}}
\begin{figure}[t]
\centering
\includegraphics[width = \columnwidth]{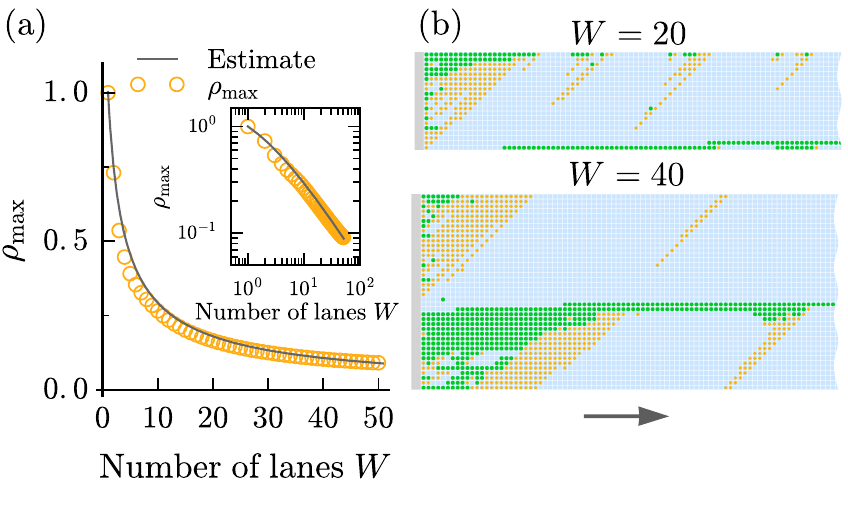}
\caption{\textbf{Dependence of the maximal density  $\rho_\mathrm{max}$ on the number of lanes $W$ for the special case $\delta=1/2$ and $L=16384$.}
 \textbf{(a)} $\rho_\mathrm{max}$ rapidly decreases with increasing numbers of lanes $W$. The decrease can be estimated by a maximum entropy argument leading to $\rho_\text{max}\left(W\right) \approx \text{Hm}(W)/W$ (solid line). \textbf{(b)} Snapshots of particle distributions in systems with $W=20$ and $W=40$ to illustrate the maximal density. Although the system is almost fully occupied at the left lattice end, the degree of occupation successively decreases and converges to the maximal density. Larger numbers of lanes exhibit more inaccessible sites and therefore a lower maximal density. In (b)  the same symbols as in Fig.~\ref{fig:model} are used to plot the system and particles. Only the leftmost part of the system is shown for illustrative purposes.  }
\label{fig:RhoMaxVsLanes}
\end{figure}

The first important question to address is how the density of a jammed system $\rho_\mathrm{max}$ is influenced by the number of lanes $W$. Our analytic approach for $W=2$ lanes, however, is not feasible for large numbers of lanes as the complexity of the underlying mathematical problem increases rapidly. Fig.~\ref{fig:RhoMaxVsLanes} shows numeric results for the maximal system density obtained from stochastic simulations of systems with up to $W=50$ lanes. Strikingly, the maximal density $\rho_\mathrm{max}$ decreases rapidly with increasing numbers of lanes. Therefore, the overall impact of topological hindrance grows accordingly and becomes the major determinant of system's dynamics. This is of importance as large numbers of lanes are often encountered in biological contexts. For example, microtubules are typically composed of 13 lanes, where our model already exhibits a maximal density of $\rho_\mathrm{max} \approx 0.2$. Let us also emphasize that, in analogy to the two-lane system, current reduction and, likewise, jamming due to topological hindrance practically saturate at low fractions of the second species, see Appendix~\ref{ap:current_density_13_lanes}.

To improve our conceptional understanding of the decrease of the maximal density with increasing numbers of lanes, we can estimate the jamming density $\rho_\mathrm{max}(W)$.
To this end, we consider groups of lattice sites with equal site index $\mu$ but different lane index $i$, which corresponds to a column of our two-dimensional lattice. Those columns can be occupied by one up to $W$ particles. The respective probabilities in a jammed system are unknown. A rough estimate is given by a maximum entropy approach  which assumes that all numbers (up to $W$) are equally likely to occur. Then, the maximal density for a system with $W$ lanes can be estimated as
\begin{equation}
\rho_\text{max}\left(W\right)=\dfrac{1}{W}\sum_{k=1}^W\dfrac{1}{k}=\dfrac{\text{Hm}(W)}{W}\xrightarrow[]{\mathrm{1 \ll W}} \frac{\ln (W)}{W},
\end{equation}
where $\text{Hm}(W)$ denotes the $W$-th harmonic number. This scaling argument, despite its simple nature, turns out to be very accurate for the case of symmetric particle mixtures $\delta=1/2$. While simulations show that the assumption of uniformly distributed numbers of particles per column is \emph{per se}  incorrect, deviations from the uniform distribution for highly and sparsely occupied columns seem to cancel out, leading to a correct mean value. An intuitive justification for this is that, for example, a completely filled column is compatible with all configurations to its left. Nonetheless, it is very unlikely to occur because only a very restricted number of states may appear to its right. On the other hand, a column with a single particle in it can occur to the right of any other state, but is only compatible with few configurations to its left. Hence this state is very rare as well. On average, high  and low occupancies are both suppressed.  By chance, the effects balance out making the mean a good approximation.

Note here that the jamming of particles below full occupation shows strong similarities with the jamming transition observed in the Biham-Middleton-Levine model~\cite{Biham1992}. The latter is a (often deterministically described) model of two perpendicularly crossing pedestrian flows that exhibits a sharp transition from a phase with finite flow to jamming.  A rigorous mathematical treatment of this transition is, however, lacking for the Biham-Middleton-Levine model~\cite{Angel2005}.

In summary, our analysis suggests that topological hindrance becomes much more significant for large numbers of lanes, where we expect it to dominate the complete dynamics. In the next section we will use a novel analytic method to construct the complete current-density relation and, thus, to relate the impact of topological hindrance to arbitrary densities. For this theory the maximal density is also a key parameter as it incorporates the full impact of the number of lanes into the relation.

\section{A theory for topological hindrance \label{sec:current_density}}
In the following, we present an approximation method which is designed to quantify the impact of  topological hindrance 
for arbitrary densities. This will enable us to derive the current-density relation of our model that successfully takes those correlations into account that lead to the failure of mean-field arguments. To do so, we show that the current-density relation can be split into a  mean-field contribution as derived in Section~\ref{sec:tophind} and a correction term that depends on the local density. This correction term can be associated with the inaccessibility of an empty lattice site and therefore quantifies topological hindrance at a certain density.
We present a construction method for this hindrance function that, in turn, allows us to compute the current-density relation.

\subsection{The hindrance function}
First, we split the particle currents into a mean-field contribution and a correction term. The particle density is independent of the lane index~$i$, such that  we find $J^S_{\mu}=\rho^S_{\mu-1}(1-\rho_{\mu})-\text{cov}(n^S_{i-1,\mu-1},n_{i,\mu})$ and $J^T_{\mu}=\rho^T_{\mu-1}(1-\rho_{\mu})-\text{cov}(n^T_{i,\mu-1},n_{i,\mu})$. Here, $\text{cov}(n^X_{i,\mu},n^Y_{l,\nu})= \langle n^X_{i,\mu}n^Y_{l,\nu} \rangle-\langle n^X_{i,\mu}\rangle\langle n^Y_{l,\nu} \rangle $ denotes the covariance of two occupation numbers.  According to the master equation, Eqs.~\ref{eq:master_eq} and \ref{eq:master_eq_bs}, the  particle currents $J^S$ and $J^T$  must be conserved on each lane in the stationary state.  In addition, the current must be independent of the lane index $i$ due to the  rotational invariance imposed by the system's cylindrical symmetry.  Hence, the total current $J=J^S+J^T$ is conserved within the whole system:
\begin{eqnarray}
J&=&\rho_{\mu-1}(1-\rho_{\mu})-\text{cov}(n^S_{i-1,\mu-1},n_{i,\mu})-\text{cov}(n^T_{i,\mu-1},n_{i,\mu}) \nonumber \\
&=&\text{const.} ,
\end{eqnarray}
for arbitrary lanes $i\in \lbrace 1, \ldots , W \rbrace$. The first summand is identical to the mean-field current derived in Section~\ref{sec:tophind}, whereas the covariances account for correlations.
We define the \emph{hindrance function} $H$ at site $\mu$ as
\begin{equation}
H_{\mu}:=\dfrac{\text{cov}(n^S_{i-1,\mu-1},n_{i,\mu})+\text{cov}(n^T_{i,\mu-1},n_{i,\mu})}{\rho_{\mu-1}}.\label{eq:hindrance_def}
\end{equation}
Then, using definition \ref{eq:hindrance_def}, the current can be rewritten as 
\begin{equation}
J=\rho_{\mu-1}\left( 1-H_{\mu} -\rho_{\mu} \right).\label{eq:current_density_discrete}
\end{equation}
Note that no approximations have been made so far. Only symmetries and conservation laws have been employed.
Considering the current-density relation of the TASEP,  $J=\rho(1-\rho)$, we find that strict accessibility of empty lattice sites is reflected by the $1$ in the second factor. In contrast, Eq.~\ref{eq:current_density_discrete} reveals that in our model an empty lattice site can be accessed only with probability $1-H_{\mu}$.
Thus, the hindrance function $H$ can be viewed as a correction to the accessibility of empty  lattice sites and reflects topological hindrance in the system. 

As the current-density relation, Eq.~\ref{eq:current_density_discrete}, defines an unclosed set of difference equations, we have to employ approximation methods to proceed with our analysis.
Often, moment closure techniques are applied to get an approximation for higher moments and thus the covariances. This would, in our case,  also fix the hindrance function which depends on them. However, based on our discussion of topological hindrance and the maximal density in Sections~\ref{sec:tophind}, we expect correlations to be  long-ranged and not constrained to certain subsegments of the lattice. Therefore, such methods do not seem promising, especially for large numbers of lanes. Instead we derive a theory similar to the considerations of Lighthill and Whitham~\cite{Lighthill1955,Blythe2007} for our model based on Eq.~\ref{eq:current_density_discrete}.
Specifically, we consider the limit of large system lengths $L\gg1$. Then, we can replace the discrete site index $\mu$ by a continuous spatial variable $x:=\mu/L\in \left( 0,1\right]$. We perform a Taylor expansion in $x$, which we truncate at first order in the lattice spacing $\epsilon:=1/L \ll 1$.  This is justified for small spatial variations of the density profile~\footnote{Due to to Eq.~\ref{eq:current_density_discrete} and the spatially constant current, spatial variations of the hindrance function are directly related to those of the density profile.}. Indeed, our stochastic simulations confirm this assumption for large aspect rations $L/W$, see Fig.~\ref{fig:pattern} (b) and Section~\ref{subsec:pattern}.
Taking everything together, the result is a  current-density relation which depends on the local density, but not on its gradient
\begin{equation}
J=\rho(x)\left[ 1-H(x) -\rho(x) \right].
\end{equation}
The current is spatially conserved (and therefore not explicitly depending on $x$) and hence  the dependence of the hindrance function $H$ on the lattice position must be implicit via the local density $\rho$.  Thus, our current-density relation can be written as
\begin{equation}
J=\rho \left[ 1-H\left(\rho \right)-\rho \right]. \label{eq:current_density}
\end{equation}
At this point, the explicit dependence of $H$ on $\rho$ is still unknown.   To make progress, we derive an approximating  function for the hindrance function $H$ that captures its physical properties. This function can then be used to predict the current \footnote{In principle, the current itself can be used for an expansion in the density but since we are specifically interested in the influence of topological hindrance the hindrance function $H$ is chosen.}.

\subsection{Constructing the current-density relation}
Since the  current-density relation of the TASEP and our model show similar curve characteristics, we  expect their qualitative phase behavior to be comparable due to the extremal current principle. We split the approximating function for topological hindrance $H$ into two corresponding regimes: $H_{\text{LD}}$ is associated with the monotonically increasing part of the current-density relation and  $H_{\text{HD}}$  with its monotonically decreasing part. These two functions can be expanded independently around extremal density values for which the behavior of $H$ can be derived using mean-field arguments. Each extremal case isolates one central aspect of transport limitation: First, the lack of particles for $\rho\rightarrow 0$, and second, jamming in case of $\rho\rightarrow \rho_\text{max}$. In the intermediate regime both transport limiting factors come into play. 

For vanishing densities  $\rho\rightarrow 0$ particles do not interact and are thus uncorrelated. In this limit we do not expect any topological hindrance and therefore the hindrance function $H$ vanishes.
Furthermore, we can  compute the derivative of $H$ by calculating the change in current due to the addition of  particles in the mean-field approximation  (see Appendix~\ref{ap:derivatives_of_hindracne}).
These two conditions allows us to determine the coefficients of an expansion of the function $H_\text{LD}$ up to linear order in $\rho$:
\begin{subequations}
\begin{align}
H_{\text{LD}}\left(0\right )&=0 ,\\
 \dfrac{\text{d}}{\text{d}\rho}H_{\text{LD}}\left(\rho\right) \biggr\vert_{\rho=0}&=\delta(1-\delta).
\end{align}
\end{subequations}
Interestingly, the result is independent of the number of lanes $W$ and depends on the species fraction $\delta$ in the simplest non-trivial way that fulfills the species exchange symmetry $\delta \rightarrow 1-\delta$.

In contrast to the low-density limit, particles are totally correlated for  $\rho\rightarrow \rho_\text{max}$. A particle can only be located at a certain lattice site if the respective target site is occupied. For this limit, we can again find the corresponding derivative of the hindrance function using mean-field arguments (see Appendix~\ref{ap:derivatives_of_hindracne}). $H\left(\rho_\mathrm{max}\right)$ is known from the definition of the maximal density $\rho_{\text{max}}$ itself. Again, we can determine the coefficients of an expansion up to linear order:
\begin{subequations}
\begin{align}
1-H_{\text{HD}}\left(\rho_\mathrm{max}\right)&\overset{!}{=} \rho_\mathrm{max} ,\\
\dfrac{\text{d}}{\text{d}\rho}H_{\text{HD}}\left(\rho\right) \biggr\vert_{\rho=\rho_\mathrm{max}}&=-\delta(1-\delta)\rho_\mathrm{max} .
\end{align}
\end{subequations}
Since the maximal density depends on the number of lanes, as discussed in Section~\ref{sec:tophind}, the same holds true for the high-density approximation of the hindrance function.  This is a major difference compared to the low-density limit. However, the low- and high-density components both show the simple dependence on $\delta(1-\delta)$. 

\begin{figure}[t]
\centering
\includegraphics[width = \columnwidth]{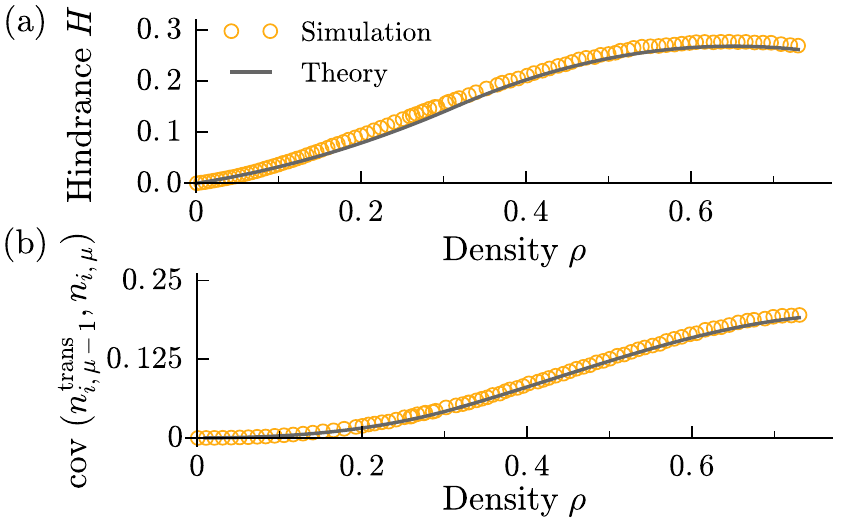}
\caption{\textbf{Comparison of the developed theory for the strength of topological hindrance with simulations for  $W=2$, $L=4096$ and $\delta=1/2$.} \textbf{(a)}~Hindrance function $H$ in the bulk as a function of the density $\rho$.  Results from simulations (symbols) verify the continuous approximation of the hindrance function (lines).  \textbf{(b)}~The construction of $H$ can be used for an accurate prediction of correlations of occupations between  neighboring sites in the bulk.  The prediction for the covariance $\mathrm{cov}(n_{i,\mu-1}^\mathrm{trans},n_{i,\mu})$ in the bulk agrees well with simulation results.  Here, $n_{i,\mu}^\mathrm{trans}:= n_{i,\mu}^T +  n_{i-1,\mu}^S$ is the total occupation of $S$ and $T$ particles of two neighboring lattice sites in transversal direction.  For the theory, $\rho_\mathrm{max}$ was taken from our analytic approach, see Appendix~\ref{ap:rho_max}.}
\label{fig:H}
\end{figure}

A closer look reveals that  linear approximations for the low- and high-density regimes do not, in general,  intersect on the interval $\left[ 0,\rho_\mathrm{max} \right]$. This would result in a discontinuous current-density relation that is unreasonable. We conclude that an expansion up to at least second order in $\rho$ is necessary. Therefore, we have to look for further physical conditions to uniquely determine the additional coefficients of the expansion.
This can be achieved by imposing a differentiable transition between the high-density and low-density regime.
At this intersection point, neither a lack of particles nor jamming is the limiting factor for  transport and hence the transition should take place at the density $\rho_{MC}$ which corresponds to the maximal current. 
Furthermore, both derivatives must vanish at $\rho_{MC}$. 
This can be translated into three conditions:
\begin{subequations}
\begin{align}
&H_{\text{LD}}\left( \rho_\mathrm{MC}\right)=H_{\text{HD}}\left( \rho_\mathrm{MC}\right),\\
&\dfrac{\text{d}}{\text{d}\rho} \rho\left(1-H_{\text{LD}}\left(\rho\right) -\rho \right) \biggr\vert_{\rho=\rho_\mathrm{MC}}=0,\\
&\dfrac{\text{d}}{\text{d}\rho} \rho\left(1-H_{\text{HD}}\left(\rho\right) -\rho \right) \biggr\vert_{\rho=\rho_\mathrm{MC}}=0.
\end{align}
\end{subequations}
Combining all properties derived for the hindrance function, we can uniquely approximate  $H_{\text{LD}}$ and $H_{\text{HD}}$ up to second order in the bulk density \footnote{In general, we may expect the current-density relation to be smooth which leads to additional constraints. However, higher order expansions would not allow one to obtain a closed analytic solution for the expansion coefficients. Since a second order approximation of the hindrance function is sufficient to find a meaningful result, we do not provide numeric approximations for higher orders.}. 
\begin{figure}[t]
\centering
\includegraphics[width = \columnwidth]{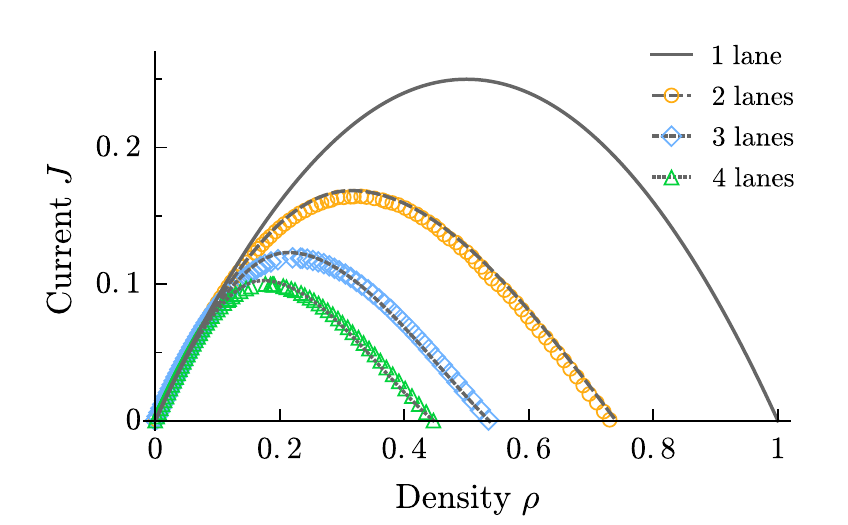}
\caption{\textbf{Current-density relation for different numbers of lanes $W$ with $L=16384$ and  $\delta=1/2$.} The theory for the current-density relation (dashed lines) agrees very well with simulation data (symbols). Although deviations from the TASEP current-density relation (solid line) increase with growing number of lanes the three regimes of the current-density relation are preserved: Low-density regime (positive derivative with respect to $\rho$), high-density regime (negative derivative) and a unique maximum. Note that the theory depends on $\rho_\mathrm{max}$ which in turn depends on $W$ and $\delta$. For the theoretical lines a fit-free approach based on the analytic result for $\rho_{\text{max}}$ (see Appendix~\ref{ap:rho_max}) was employed for two lanes. For three and four lanes results where taken from the simulations to ensure that deviations originate from the theory itself not the estimate of $\rho_{\text{max}}$. The particle current refers to an arbitrary lane of the system in order to allow for a comparison to the TASEP. \label{fig:current_density_multilane}}
\end{figure}
The resulting hindrance function is shown in Fig.~\ref{fig:H} (a) for a system with two lanes and species fraction $\delta=1/2$; The corresponding current-density relations for multiple numbers of lanes $W$ are shown in Fig.~\ref{fig:current_density_multilane}. As can be seen, the results of our approximations are in very good agreement with the stochastic simulations and capture the main characteristics of the current-density relation. At this point, let us also emphasize that the theory can be applied for general $\delta$ and $W$. It does not involve any explicit dependence on the number of lanes $W$ but depends on $\rho_\mathrm{max}$ which, in turn, depends on $W$ and also on $\delta$. Therefore, $\rho_\mathrm{max}$ is a key quantity in our theory, which determines the current-density relation and hindrance function and thus quantifies topological hindrance in general.

Since the current is very well  approximated  with this method, the exact relation given by Eq.~\ref{eq:hindrance_def} allows us to infer the underlying correlations. This leads to a prediction for the covariances,
\begin{equation}
\text{cov}(n^S_{i-1,\mu-1}+n^T_{i,\mu-1},n_{i,\mu})\approx  \rho H\left(\delta,W;\rho\right),\label{eq:cov_vs_H}
\end{equation}
where we have accounted explicitly for the dependence of the hindrance function on the control parameters $W$ and $\delta$. 
This prediction of the covariances is confirmed by the stochastic simulation shown in Fig.~\ref{fig:H} (b), as expected from the method's accurate prediction of the current. 

In summary, our analysis shows that modifications of the current-density relation in our system are caused by particle correlations. These correlations are accounted for by a hindrance function $H$, which quantifies the degree of inaccessibility of empty lattice sites, and therefore quantitatively characterizes topological hindrance.
We construct the hindrance function in terms of a gradient expansion. This expansion is determined by physical constraints with respect to the high and low-density limits as well as the transition to the  maximal-current phase.
Numerical simulations validate our method.
\subsection{The phase behavior of collective two-species transport}
So far, we have computed the relation between the average density and particle current but haven't addressed how these observables are connected to the system's control parameters.
In this section we determine the response of current and density to a change in the control parameters $\alpha$ and $\beta$ and thereby derive the complete phase diagram.
As mentioned above, the extremal current principle implies the existence of the three phases as in the TASEP (low-density, high-density, and maximal-current phase), due to the similarity in the form of the current-density relation.
Then, for various driven lattice gas systems, current conservation at the respective boundary is used to relate the system's current and density to in and out rates. 
Due to non-trivial correlations in our system, it is, however, not possible to simply treat the boundaries as reservoirs with density $\rho_L = \alpha$ and $\rho_R=1-\beta$ as is done for the TASEP. In Appendix~\ref{ap:phase_diagram} we present ways to account for correlations at the edges of the system. This involves determining the response of current and density to the in rate $\alpha$ in the low-density phase and to the out rate $\beta$ in the high-density phase. 
 
\begin{figure}[t]
\centering
\includegraphics[width = \columnwidth]{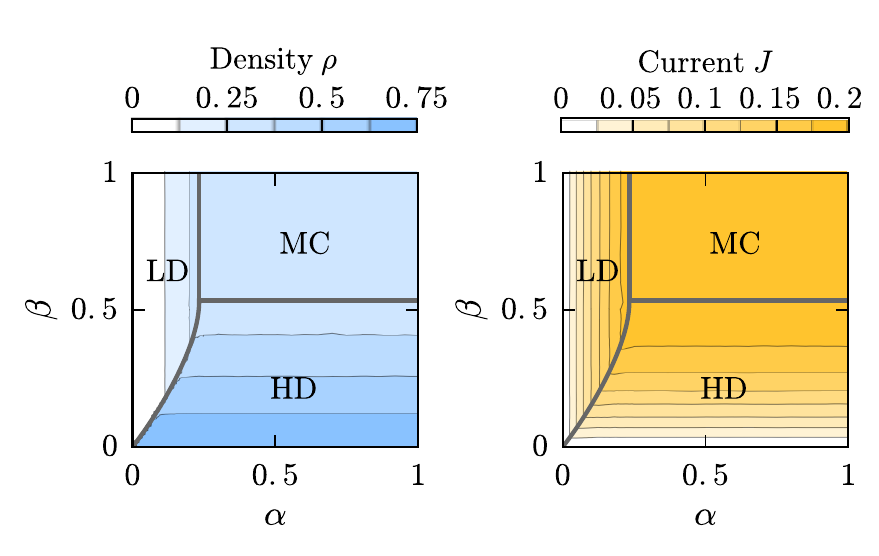}
\caption{\textbf{Phase diagram for the special case $W=2$, $L=4096$ and $\delta=1/2$.} Color denotes density (left panel, blue) and current (right panel, orange) obtained from stochastic simulations. Three different phases can be identified: The low-density phase (LD) where current and density only depend on the in rate.  The high-density phase (HD) where current and density only depend on the out rate. And the maximal-current  phase (MC) that is independent of both boundary parameters. The theory (bold lines) accurately predicts the phase boundaries and is valid for general $\delta$ and $W$ as soon as the maximal density is known. Here, $\rho_\mathrm{max}$ was taken from the analytic approach, see Appendix~\ref{ap:rho_max}.}
\label{fig:phase}
\end{figure}

The low-density to maximal-current (LD-MC) and the high-density to maximal-current  (HD-MC) transitions take place when the respective current matches the maximal bulk current. The latter can be computed with the methods outlined in the previous section.  Analogously, the low-density to high-density (LD-HD) transition takes place when both currents are identical. The resulting phase diagram for the special case of two lanes and a species fraction of $\delta=1/2$ is shown in Fig.~\ref{fig:phase}. The theoretical boundaries  were calculated based on the analytically determined value for the maximal density for this choice of parameters (see Appendix~\ref{ap:rho_max}) and hence are derived without any free parameter. The results agree very well with the data obtained from stochastic simulations.
Changing $\delta$ does not affect the existence of phases but interpolates between the boundaries known for the TASEP and those presented in Fig.~\ref{fig:phase}. Increasing the number of lanes $W$ lowers the maximal density (see Fig.~\ref{fig:RhoMaxVsLanes}) and increases the parameter range of the maximal-current  phase, while the corresponding current is reduced. Hence, as predicted by the extremal current principle, given the number of lanes and species ratio, the phase diagram is  topologically equivalent to the one obtained for the normal TASEP~\footnote{There is actually an additional phase at an out rate of strictly zero. This phase has no counterpart in the TASEP and is discussed in detail in Appendix~\ref{ap:rho_max}}.  Note that our theory holds for general values of the system width $W$ and the species fraction $\delta$. 

\subsection{Non-equilibrium pattern formation \label{subsec:pattern}}

\begin{figure}[!tb]
\centering
\includegraphics[width = \columnwidth]{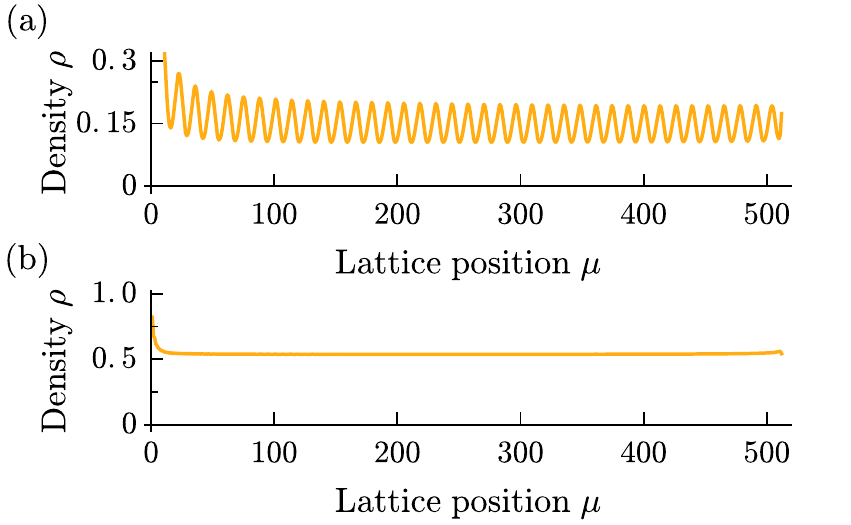}
\caption{\textbf{Density profiles $\rho_\mu$ for large and small system widths $W$ compared to the length $L=512$.} \textbf{(a)} The stationary density distribution for $W=13$ lanes shows an oscillatory pattern with a wavelength equal to the number of lanes.  The system exhibits a constant current despite the density oscillations.   \textbf{(b)} Stationary density distribution for $W=2$ lanes. The system exhibits a flat density profile within the bulk. Other parameter values: $\alpha =0.6$, $\beta=0.2$ and $\delta =1/2$ in both simulations.}
\label{fig:pattern}
\end{figure}

The construction of the hindrance function shown above relies on an expansion of the density $\rho$ in terms of the spatial variable $x=\mu\ /L$. Specifically, we assumed that gradients in the density are negligible, which is justified for slowly varying density profiles. While this assumption as illustrated in Fig.~\ref{fig:pattern} (b) leads to accurate results in most cases, we observe surprising exceptions to this behavior for systems with a small aspect ratio $L/W$: If we sufficiently decrease the system length $L$ or increase the system width $W$, our stochastic simulations reveal spatial oscillations of the stationary average particle distribution. As shown in Fig.~\ref{fig:pattern} (a), the density $\rho_{i,\mu}=\langle n_{i,\mu} \rangle$ oscillates along the longitudinal direction (Greek index $\mu$) with a wavelength equal to the width of the system $W$. Due to rotational symmetry these wave-like patterns of the density profile are equally present for each lane $i$. Further, oscillations are sustained for thousands of lattice sites almost without any decay and therefore show remarkable robustness. This behavior is in stark contrast to the TASEP, which (except for boundary layers) exhibits a constant density profile. In general, pattern formation is rarely observed  in lattice gas models and has so far been found predominantly in the form of segregation or localization effects~\cite{Johann2014,Reithmann2016,Graf2017,Pinkoviezky2017}.

As stated above, the continuous approximation Eq.~\ref{eq:current_density} is by virtue of its construction incapable of describing these varying density profiles. Also, the species current fraction $\delta$ may in this case significantly differ from the species density fraction, see SAppendix~\ref{ap:species_fraction}.
It is, however, worth noting that our theory still provides a good approximation for the current-density relation in terms of the \emph{total} system density (i.e. when averaging over an oscillating density profile) for symmetric species fraction $\delta=1/2$.
To describe the density profile itself, it is necessary to include a dependence not only of the local density, but also its spatial change in the hindrance function $H$. The problem is comparable to the boundary layers of the TASEP, which can not be captured by a first order continuous theory, but are predicted by exact or higher order solutions~\cite{Lakatos2006,Evans2011,Chou2011b}. In future work, it would be interesting to address if such a construction or some alternative analytic method is capable of reproduce these intriguing patterns.

\section{Robustness and biological relevance}
\label{sec::biological_relevance}
\begin{figure}[!tb]
\centering
\includegraphics[width = \columnwidth]{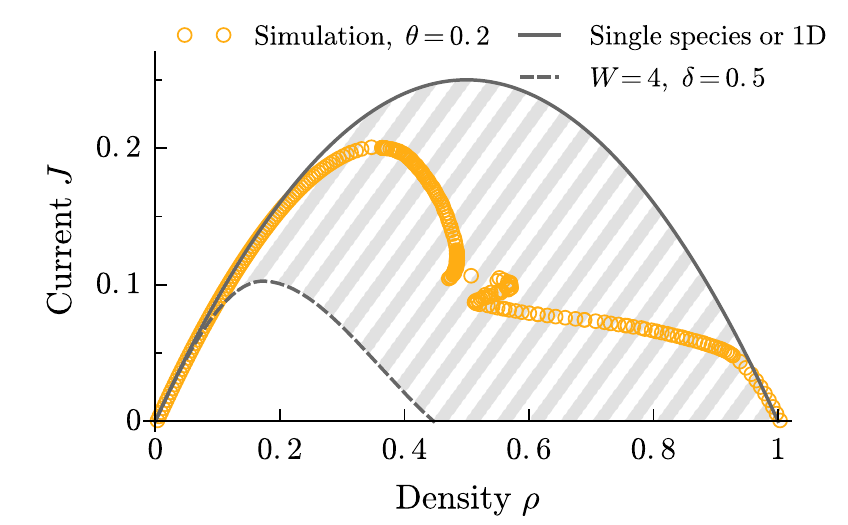}
\caption{\textbf{Current-density relation of a model with non-strict lane hopping on a system composed of $W=4$ lanes.} Simulation results (symbols) for the current and density of a model where the lane-switching species switches the lane only with probability $\theta=0.2$ in each step; With probability $1-\theta$ the second species follows its current lane. Topological hindrance significantly suppresses current and density also for this model, where lane switching occurs seldom. The resulting data points are in the area between the current-density relation  of the single-species case or one dimensional model (solid line) and the one of our model (dashed line). For models similar to the one presented in this work but with more realistic particle dynamics (e.g. random particle attachment/detachment, non-strict lane hopping, dimeric particles) we expect current suppression due to topological hindrance to be stronger for parameters that lead to bigger differences in both curves (shaded area). Thus, our theory provides qualitative insights also in more complex models.  Note that the simulations suggest a non-injective current-density relation  at the onset of jamming of the original model.  Other parameter values: $\delta=1/2$, $L=16384$.}
\label{fig:non-strict_lane_hopping}
\end{figure} 

To establish a generic theory for two-species transport and to enable an analytic study of topological hindrance, we previously made several assumptions which do not necessarily hold in a physical or biological context. To probe the robustness of our results  and the relevance of topological hindrance  further we now turn to extended versions of our model that account for biomolecular features of molecular motors. 

Firstly,  kinesins are likely to not strictly follow a unique pathway, i.e. lane switching does not occur at every step but stochastically. Adding this modification to our model  mitigates effects such as a strict maximal density and phase transitions since any  inaccessible site of the original model can now be accessed  in principle on long time scales.
Nonetheless, also in this case  topological hindrance is of significant impact. To demonstrate this, we performed simulations with a four-lane system in which lane-switching particles switch lanes with a probability of $\theta=0.2$ and track their respective lane otherwise. The results are shown in Fig.~\ref{fig:non-strict_lane_hopping}.  While this system does not strictly follow the current-density relation as predicted by our theory, it demonstrates that topological hindrance has a strong influence on collective transport as particle the current is heavily suppressed: For example, the average particle current is reduced by more than 50\% at densities around the maximal density of the original model. For higher lane numbers current reduction is even more pronounced. This shows that topological hindrance can have substantial influence also on systems where particles switch lanes rarely but that are composed of many lanes.

\begin{figure}[!tb]
\centering
\includegraphics[width = \columnwidth]{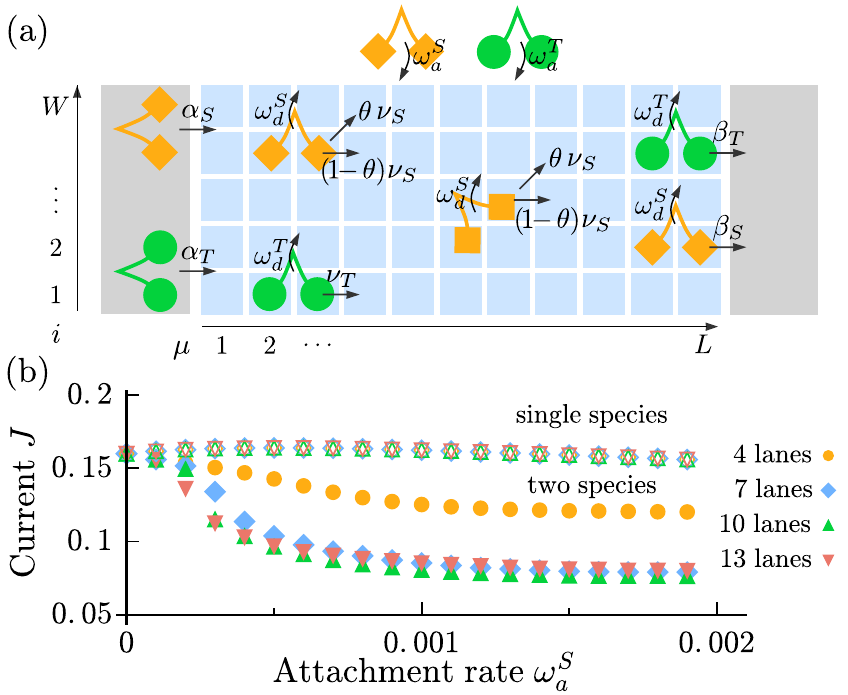}
\caption{\textbf{Illustration of the extended model and current reduction for different numbers of lanes.} 
\textbf{(a)}~ Dimers populate a lattice of width $W$ and length $L$. Periodic boundary conditions are employed along the transversal direction. Dimers of species T move their rear leg to right neighboring lattice site of the front leg at rate $\nu_T$. Dimers of species S move their rear leg to right neighboring lattice site of the front leg at rate $(1-\theta)\nu_S$ and to the upper-right neighboring lattice site of the front leg at rate $\theta \nu_S$. All particles exclude each other.  Dimers enter the system either at the left boundary at  rates $\alpha_S$ and $\alpha_T$ or randomly attach to the lattice at respective rates $\omega^S_a$ and $\omega^T_a$. Lane switching particles may either attach with both legs on the same lane at rate $(1-\theta)\omega^S_a$ or, alternatively, with feet on neighboring lanes at rate $\theta \omega^S_a$. Particles exit the system at the right boundary at rates $\beta_S$ and $\beta_T$ or stochastically detach from the lattice at rates $\omega^S_d$ and $\omega^T_d$. \textbf{(b)}~Simulation results show the current for systems with different numbers of lanes $W$ without any spiraling motion (empty symbols) and with rare stochastic lane switching (filled symbols) of $\theta=0.1$. If no lane switching is possible the current is not affected by the number of lanes and reduces to the single-species TASEP with Langmuir kinetics and dimeric particles~\cite{Parmeggiani2003,Parmeggiani2004,Pierobon2006}  as identical hopping rates for both species were chosen.  However, if the second species is allowed to switch lanes a rapid current reduction similar to the one described above for the main model can be observed.  Note that the current is no longer conserved within the system such that a spatial average over a lane was taken. Other parameter values: $\omega_d^S=\omega_d^T= \omega_a^T = 0.001$, $\nu_S=\nu_T=1$, $\alpha=0$, $\beta=0.5$, $L=2048$.}
\label{fig:bio_model}
\end{figure}

To challenge our results further, we performed stochastic simulations of an extended model that accounts for other important biomolecular features of molecular motors: In addition to the stochastic lanes switching described above, particles may attach (rate $\omega_a^X$) and detach (rate $\omega_d^X$) randomly along the lattice (Langmuir kinetics) \cite{Parmeggiani2003,Parmeggiani2004} and occupy two lattice sites to account for a dimeric structure \cite{Pierobon2006} of most molecular motors (see illustration in Fig.~\ref{fig:bio_model}). In analogy to our  analysis of the minimal model shown in Fig.~\ref{fig:current_denisty}, we increase the particle attachment rate $\omega_a^S$ of the spiraling motor species while keeping the attachment rate of the tracking species $\omega^T_a$ constant. An example for the corresponding average particle current per lane is depicted in Fig.~\ref{fig:bio_model}. Even in this greatly expanded model topological hindrance causes a significant decrease in the particle current already at low fractions of the second species. In particular, the impact of topological hindrance follows the same principles as discussed in section~\ref{sec:tophind}: It increases rapidly with an increasing number of lanes and it sets at already small fractions of a second species.
Taking all results together, we conclude that topological hindrance most likely is of relevance also in biological contexts, and the theory described here allows one to estimate its impact.
Another interesting question in this context is, whether a phenomenon similar to the density patterns which we  observe in the main model exists in a biological context. As this strongly depends on the system which is considered we can, at this stage, not provide a complete analysis. Nevertheless, to give an outlook  on the impact of model extensions on pattern formation a brief discussion is given in Appendix~\ref{ap:robustness}. 

Moreover, we would like add that random particle attachment and detachment may, in general, modify our results. In particular, we expect that adding Langmuir kinetics to our model leads to the emergence of additional phases~\cite{Parmeggiani2003,Parmeggiani2004} such that our results --while providing a mathematical foundation---can not be transferred directly~\footnote{For example the dynamics of the system may be dominated by the bulk instead of just either of the two boundaries}.  Whereas  a complete analysis of the jamming behavior is out of scope of this work,  the analysis of our fully extended model (Fig.~\ref{fig:bio_model}) strongly suggests that whenever transport occurs over significant length scales, topological hindrance can be expected to significantly impact  the particle current in general. 

To provide a further link from our theoretical work to biological systems we finally would like to address how our results could be tested experimentally. Two central predictions our model makes in this context are (1) current reduction and (2) periodic particle jams:
(1) The average particle current in system with two species is expected to be significantly smaller compared to a single species system (see Fig.~\ref{fig:bio_model}). This could be measured experimentally using time-resolved single molecule techniques. (2) Topological hindrance is related to spatial correlations in the particle arrangement and movement which may lead to patterns in the particle density. These could be resolved experimentally using techniques  described previously for example by Maurer et.~\textit{al.} \cite{Maurer2014}.
One hurdle to overcome is to construct suitable TIRF setups which do not obstruct the helical pathways of the spiraling motor species, similar to those described in references \cite{Can2014,Bugiel2018}.

\section{Summary and Conclusion\label{sec:conclusion}}

In this work, we have studied collective transport of two particle species with distinct gaits and therefore different directions of motion on a cylinder. 
As a key result, we found that the presence of different stepping modes enhances steric interactions and compromises particle dynamics in ways not seen in single-species systems. This additional hindrance changes the macroscopic transport behavior in a way that can -- to the best of our knowledge -- not be accounted for by previous models or analyses.
The combination of a lane-tracking and a lane-switching species gives rise to unexpected phenomenology in the following fashion: 
\begin{enumerate}
\item Systems with a mixed population of both species always jam at densities far below full occupation.
\item Closely related to that, a system with a mixed population is always characterized by a lower average particle current at a given total system density.
\item The proportion of the second species present has a crucial influence on this behavior: Current reduction sharply sets in at very small fractions of the second species and practically saturates for fractions larger than  $\delta \approx 5\%$. 
\item In contrast to single-species transport, the system width $W$ is an important determinant of collective dynamics. Increasing $W$ rapidly decreases the density at which particles jam and also rapidly increases the degree of current reduction. 
\item The average particle distribution (density profile) shows wave-like patterns for small aspect ratios $L/W$ of the system. This again contrasts single-species transport and many transport models, as those typically exhibit a spatially constant or linear density profile.
\end{enumerate}

The above observations can be traced back to the following microscopic origin: While for transport with a single species each position can be equally accessed and vacated from a certain number of directions, our model shows a more intricate connectivity. A single lattice site may now be accessed from two different directions but can be vacated in one direction only. This simple change in the network topology has substantial influence on collective behavior. It creates inaccessible lattice sites and slows down particle motion due to intrinsic bottlenecks  at points where trajectories intersect. This concept, that steric interactions are amplified by the network topology, forms the basis of our theoretical approaches, which sheds light on the following issues.
\begin{enumerate}
\item We provide a detailed analysis of the jamming process. The decrease of the jamming density with increasing system width $W$ is well approximated by $\rho_\mathrm{max} \rightarrow {\ln (W)}/{W}$ for large $W$ and symmetric species mixtures $\delta=1/2$. This explains the relevance of the system width $W$ for topological hindrance and suggests convergence to a vanishing jamming density for large system widths. We specifically compute the jamming density in the case $\delta=1/2$ and $W=2$, which validates our understanding of the jamming process.

\item Even at densities well below the jamming transition many lattice sites are inaccessible. We show that  inaccessibility is quantified by a hindrance function which naturally arises from particle correlations.  In doing so, we obtain the current-density relation for arbitrary species fractions and system widths.
This method is not restricted to our specific system. It reproduces the correct current-density relation of the TASEP and we expect it to be applicable to other TASEP-like systems that exhibit cylindrical symmetry.
 
\item The current-density relation is used to compute the phase diagram of our model. In this way, we globally relate central macroscopic observables, namely the average particle current and distribution, to the model's control parameters, the particle in and out rates. 
\end{enumerate}

An important conclusion of our study is that the reduction to one dimension is \emph{a priori} not possible for collective transport of differently moving agents along a cylinder. 
Although dynamics for each species alone are effectively one-dimensional and are successfully described by mean-field methods, the joint system deviates from this behavior.
Our analysis overcomes this limitation and establishes a framework for mixed populations comprising distinct particle species. It further identifies topologically amplified hindrance as a central determinant in this context.
The derivation of the hindrance function not only quantifies the strength of topologically amplified hindrance but also implements an \emph{effective} mapping to a one-dimensional system. This mapping is useful two ways. First, it reduces the complexity of the mathematical problem by integrating out the transversal spatial dimension. Second, it also allows one to infer possible microscopic interactions whenever only a one-dimensional projection is visible. For example, in a biological context it is still technically challenging to resolve the motion of molecular motors below the diameter of their respective filament. Hence, the projection of particle positions on the filament contour is the standard information accessible in experiments. This holds particularly true in experimental setups with a multitude of moving particles, where super-resolution techniques are not always feasible. As our work shows, it is a highly non-trivial task to relate macroscopic information back to individual microscopic interactions, and our theory might yield valuable insights in this respect. 
But our considerations also raise the question of how much actual biological systems, such as molecular transport along microtubules, are affected by topological hindrance. While some of our specific results (e.g. a strict maximal density or phase transitions) are weakened by processes like random particle attachment/detachment and non-strict lane hopping, we still expect topologically amplified interactions to play a critical role. Our analysis regarding model modifications  in Section~\ref{sec::biological_relevance} suggests that, whenever transport occurs over significant length and time scales, topological hindrance most likely changes collective behavior substantially.  An important conclusion in the biological context is that particle jamming due to topological hindrance takes place at significantly lower particle densities as compared to single species transport: Fig.~\ref{fig:bio_model} suggests that even slight deviations from the straight stepping behavior (10\% in this case) reduce the current to half of the value  as compared to the single species case at identical cytosolic  particle concentrations. Thus, jamming transitions might be more relevant also \textit{in-vivo} than thought previously.  Therefore, it might be revealing to probe this phenomenon experimentally in multi-species setups and to speculate about its  biological implications.

In these ways, our work has various consequences for intracellular transport and yields a generic theory for collective motion of differently moving agents. On a theoretical level, our analysis is in line with results of the TASEP for single-species transport. But since the TASEP has relevance beyond the field of transport processes,  our results form also a bridge to other intensively studied fields of statistical physics: Our model constitutes a simple system far from thermal equilibrium that can be investigated in great detail with the theory developed in this manuscript. In that, it exhibits a wide range of phenomena that are insufficiently understood and have rarely been observed in analytically well accessible models such as driven lattice gases. First, in contrast to many other driven lattice gas models, the average particle distribution exhibits complex pattern formation. While a detailed mathematical study of these patterns is beyond the scope of this work, our construction of the hindrance function and the corresponding reduction to one dimension can form a basis for a more rigorous investigation. Second, our model features a non-trivial jamming transition. While jamming has severe and intricate implications also for the TASEP and similar models, the origin of jams in our model is not only the consequence of mere overcrowding, but also of the spatial arrangement of the constituents. This is of interest as jamming transitions arise in many different problems such as traffic flow, granular media, or glassforming liquids~\cite{LiuBook2001}. Nonetheless, it is still a demanding and ongoing task to establish a concise theoretical framework for these phenomena~\cite{Schall2012,Liu2014},  although it is an open question whether a general connection exists~\cite{LiuReview2010,LiuReview1998}. The challenge involved in finding a mathematical formulation of jamming processes also becomes evident in view of the Biham-Middleton-Levine model~\cite{Biham1992} for traffic flow: Although the model has been extensively studied for more than 15 years, a rigorous analytic approach to describe its jamming transition is still elusive and considered to be an unsolved mathematical problem~\cite{Winkler2004,Winkler2007,Angel2005}. In this light, our study  opens a new perspective on jamming processes as well as pattern formation in terms of simple, analytically accessible models, and further elucidates the intriguing principles of collective behavior emerging in non-equilibrium systems.

\begin{acknowledgments}
We thank Silke Bergeler, Isabella Graf, Johannes Knebel, Timo Kr\"uger, Louis Reese, and Moritz Striebel for discussions and feedback to the manuscript.
This research was supported by the German Excellence Initiative via the program ``NanoSystems Initiative Munich'' (NIM), and the Deutsche Forschungsgemeinschaft (DFG) via project B02 within the Collaborative Research Center (SFB 863) ``Forces in Biomolecular Systems''. We also gratefully acknowledge financial support by the DFG
Research Training Group GRK2062 (Molecular Principles of Synthetic Biology).

P.W. and E.R. contributed equally to this work.

\end{acknowledgments}

\appendix
\section{Derivation of the maximal density $\rho_\mathrm{max}$ for $\delta=1/2, W=2$ \label{ap:rho_max}}

One of the critical differences of our model in comparison to one-species transport is complete jamming of particles at densities far below full occupation. In this section, we will focus on these completely jammed configurations, determine their number of inaccessible lattice sites, and thus the maximal density $\rho_\mathrm{max}$.

\begin{figure}[t]
\centering
\includegraphics[width = \columnwidth]{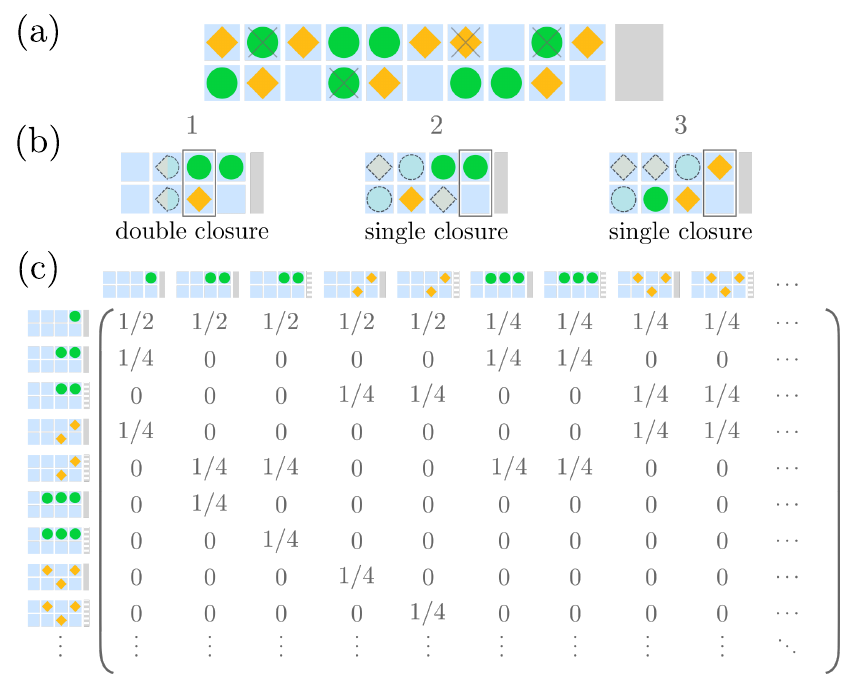}
\caption{\textbf{Illustrations for the filling process, $\beta=0$, $\alpha \rightarrow 0$.} \textbf{(a)} During the filling process loose particles (crossed particles) are created. Loose particles are particles that can be removed from a jammed configuration without allowing further particle motion. Thus, they are suppressed in any stationary state that exhibits an arbitrary non-vanishing current. \textbf{(b)} If a column is fully occupied (box in group 1), the next particle (shaded particles) can not pass this column and therefore a truncation of the state space with respect to particles to the right is possible. This is referred to as double closure. If a linear array of at least two particles of the same species is followed by a particle of the other species, either a double closure is created or a single occupied column can be identified (boxes in groups 2 and 3) beyond which none of the following particles (shaded particles) can pass. This also allows one to truncate configurations to the right of this column and is referred to as single closure. \textbf{(c)} Transition matrix of the stochastic process for  $\beta=0, \alpha \rightarrow 0$. The state space reduces to arrays of particles of the same species that follow a double (solid end) or single closure (shaded end). A single particle can only arise after a double closure as single closures immediately create arrays of length two.  In (a) and (b) a gray area denotes the system boundary.}
\label{fig:filling_illustration}
\end{figure}

Complete jamming occurs only for vanishing out rates. One might expect that it is possible to focus on the case $\beta=0$. This results in a much easier stochastic process to which we will refer as \emph{filling process}. Our stochastic simulations, however, show that a more careful analysis is required when $\beta$ vanishes: The average system density exhibits different values for arbitrarily small out rates $\beta\rightarrow 0$ than for out rates that strictly equal zero, $\beta=0$. For example, for $W=2$ and $\delta=1/2$ our simulations show that the density converges to a constant value of $\rho_\mathrm{max}\approx 0.73$ for small but non-zero values of $\beta$ (tested for different values of $\beta$ up to  $10^{-10}$). Setting $\beta=0$ the system realizes a different density of $\rho_\mathrm{filling}=0.79$. Our analysis of completely jammed configurations hence has to treat the cases $\beta=0$ and $\beta\rightarrow 0$ separately. 

To obtain a relation between the densities corresponding to $\beta=0$ and $\beta\rightarrow 0$, it is of relevance to understand how such a discontinuous behavior of the density can arise. Ultimately, it can be traced back to an instability of subconfigurations that arise in the filling process. A possible configuration is shown in Fig.~\ref{fig:filling_illustration} (a). The filling process creates configurations that contain particles which do not block other particles (crossed particles in  Fig.~\ref{fig:filling_illustration} (a)). Consequently, it is possible to remove these \emph{loose particles} from a completely jammed system without allowing further particle motion. In the stationary state, however, each particle that is removed from the system has to be compensated by a new particle on average. Therefore, loose particles are suppressed in any stationary state that exhibits a non-vanishing yet arbitrarily small, stationary current; The system undergoes a discontinuous phase transition in the density as soon as particles may leave it. 

Based on this reasoning, it is possible to derive analytic expressions of the average system density created by the filling process $\rho_\mathrm{filling}$ and the maximal (dynamic) system density $\rho_\mathrm{max}$ corresponding to $\beta \rightarrow 0$. For simplicity, we restrict our discussion to $W=2$  lanes and $\delta=1/2$ throughout this section.
Our analysis includes the following steps: We first focus on the filling process with $\beta=0$. For this stochastic process, it is possible to obtain an \emph{exact} analytic solution for the probability of finding an arbitrary configuration and therefore for $\rho_\mathrm{filling}$. The result further allows us to compute the probability $p_\mathrm{loose}$ of finding a loose particle in a configuration created by the filling process. When a non-vanishing current is established, the presence of loose particles is suppressed and the density decreases correspondingly, which will be computed in the last step of our derivation.

\subsection{The filling process}
The filling process is defined as the stochastic process arising for $\beta=0$. Typically, in jammed configurations the system properties are independent of the in rate $\alpha$, which suggests to also consider the limit $\alpha \rightarrow 0$. Then, we can assume that particles arrive at the right lattice end independently which further simplifies our discussion. Our mathematical formulation should describe the sequence of lattice occupations when particles are added sequentially. Specifically, a (discrete) time step is defined by a new particle getting stuck in the system. A na\"ive way to denote the state space would be to enumerate the $3^{2 \times L}$ possible lattice occupations. However, this leads to a very irregular and high-dimensional state space that precludes an exact solution. Fig.~\ref{fig:filling_illustration} (b) shows several configurations which are jammed at the lattice end but empty at the front. Light colored particles denote the possible positions at which the next particle can end up. In all cases, the next particle can not pass the column that is marked with a box. This means that its final position is independent of the configuration to the right of this box. We can identify such a column where particles can't pass in two different cases: First, whenever a column is filled completely (box in group 1, Fig.~\ref{fig:filling_illustration} (b)) no more changes can be made to the configuration to the right of this column. We will refer to this column as \emph{double closure}. Second, whenever a linear array of particles of the same species is followed by a particle of the other species a column that can't be passed can be identified. In this case, either a double closure may be created or a new kind of closure arises. Specifically, if no double closure is is created we can always identify a half occupied column (boxes in groups 2 and 3, Fig.~\ref{fig:filling_illustration} (b)) beyond which no further change is possible. We will refer to this column as a \emph{single closure}. The above reasoning then suggests the following truncation scheme for the state space: Whenever a closure occurs, we can truncate the state space with respect to the occupation to the right of it. This significantly reduces the complexity of the state space. The remaining possible configurations are linear arrays of a single species to the left of either of the two different types of closures. Given this reduced state space, it is possible to write down the resulting (infinite dimensional) transition matrix that characterizes the underlying stochastic process. The result reads
\bequa
M &=& \left( 
  \begin{matrix} 
    1/2 & 1/2 & 1/2 & 1/2 & 1/2 & 1/4 & 1/4 & 1/4 & 1/4 & \hdots\\
    1/4 & 0 & 0 & 0 & 0 & 1/4 & 1/4 & 0 & 0 & \hdots\\
    0 & 0 & 0 & 1/4 & 1/4 & 0 & 0 & 1/4 & 1/4 & \hdots\\
    1/4 & 0 & 0 & 1/4 & 1/4 & 0 & 0 & 1/4 & 1/4 & \hdots\\
    0 & 1/4 & 1/4 & 0 & 0 & 1/4 & 1/4 & 0 & 0 & \hdots\\
    0 & 1/4 & 0 & 0 & 0 & 0 & 0 & 0 &  0 & \hdots\\
    0 & 0 & 1/4 & 0 & 0 & 0 & 0 & 0 &  0 & \hdots\\
    0 & 0 & 0 & 1/4 & 0 & 0 & 0 & 0 &  0 & \hdots\\
    0 & 0 & 0 & 0 & 1/4 & 0 & 0 & 0 &  0 & \hdots\\
    \vdots & \vdots & \vdots & \vdots & \vdots & \vdots & \vdots & \vdots & \vdots &  \ddots \\
  \end{matrix}
\right) , \nonumber
\eequa
where the states are enumerated as shown in Fig.~\ref{fig:filling_illustration} (c). This matrix encodes the evolution of probabilities for occupations after a closure, and the frequencies at which double as well as single closures occur. For an infinite system the occupational probabilities of the stochastic process will converge to a steady state which corresponds to the eigenvector of the stochastic matrix $M$ with eigenvalue one. The result reads
\bequa
  P_\mathrm{steady\ state} &=& \left( 
  \begin{matrix} 
   2 \sqrt{3}-3 \\
   \frac{1}{4}(2-\sqrt{3}) \\
   \frac{1}{4}(13\sqrt{3} - 22) \\ 
   \frac{1}{4}(2-\sqrt{3}) \\
   \frac{1}{4}(13\sqrt{3} - 22)\\ 
   (\frac{3}{16}(59\sqrt{3} -102))^1  \\
   (\frac{1}{16}(2-\sqrt{3}))^1  \\ 
   (\frac{3}{16}(59\sqrt{3} -102))^1  \\
  ( \frac{1}{16}(2-\sqrt{3}))^1  \\
   (\frac{3}{16}(59\sqrt{3} -102))^2 \\
   (\frac{1}{16}(2-\sqrt{3}))^2 \\
   (\frac{3}{16}(59\sqrt{3} -102))^2 \\
   (\frac{1}{16}(2-\sqrt{3}))^2 \\
   \vdots \\
  \end{matrix}
   \right) .
\eequa
With these steady-state probabilities for lattice configurations  we can derive an exact result for the average particle density in an infinite system. We compute the frequencies of  blocks of subconfigurations with a closure at the left being added to the system. These frequencies can be determined by the transition rates and the steady-state probabilities. For example, an array of two $T$-particles following a double closure (second configuration in the enumeration scheme of Fig.~\ref{fig:filling_illustration} (c)) can be ``closed'' by an arriving $S$-particle such that a stable block dimension 2$\times$2 occupied by three particles is created. As the probability of an $S$-particle arriving at the respective site is $1/4$ the corresponding frequency for this block to occur is $1/4 \, P_2$. We weigh these frequencies with the block lengths and densities which yields the system density in the limit $L\rightarrow \infty$.  Our analytic approach results in a value of $\rho_\mathrm{filling} = 0.79$. Notably, we made no approximations in this computation. Consequently, it should coincide with the system densities for $\beta=0$ and $\alpha \rightarrow 0$. Indeed, simulations with respective parameter values confirm our result. 

\subsection{The maximal dynamic density $\rho_\mathrm{max}$}

As illustrated above, the presence of unstable, loose particles will lead to a different density as soon as $\beta$ is set from zero to arbitrarily small values. Therefore, computing the number of loose particles that arise in the filling process is key to determine the density decrease as soon as the system becomes dynamic. In the first step to compute the maximal density $\rho_\mathrm{max}$ we will determine the probability of finding a loose particle. In the second step we will show how to correctly replace the corresponding occupational density and thereby estimate $\rho_\mathrm{max}$.

Loose particles can only arise whenever both lanes are occupied at a given position $\mu$, i.e. within a double occupied column. The probability of double occupied columns created by the filling process are easily computed from the filling density. For further convenience, we introduce a new random variable $c_\mu \in \{0,1\}$ that equals 1 if the $\mu$-th column is fully occupied (irrespective of the particle species, i.e. for $n^X_{1,\mu}=1 \land n^X_{2,\mu}=1$) and 0 otherwise. Then,
\bequa
 P(c_\mu \!=\!1) &=& 2 (\rho_\mathrm{filling} - \frac{1}{2}),
\eequa
where $P(c_\mu \!=\!1)$ is the probability of finding a fully occupied column. Then, a loose particle occurs in a double occupied column in two different situations: First, when this column is followed by a column to its left that is occupied by only one particle. This will always result in a loose particle in the double occupied column. Second, if the double occupied column is followed by another double occupied column to its left where a particle from each species is present. Then, the two particles of the left column are directed to a single site which creates a loose particle next to this site. To compute the corresponding probabilities we use the exact solution of the filling process. In detail, we can determine the probability of a double closure being followed by a fully occupied column. This scenario is identical to the conditional probability of a double occupied column being followed by another double occupied column and, using the above solution of the filling process, is given by $p(c_\mu \!=\!1| c_{\mu-1}\!=\!1)\approx 0.64$. In half of these cases the fully occupied column on the left is populated by particles from different species. This corresponds to the second case described above and therefore creates a loose particle. If the double occupied column is \emph{not} immediately followed by another double occupied column (corresponding to $(1-p(c_\mu \!=\!1| c_{\mu-1}\!=\!1)$) the first case as described above occurs and again a loose particle is created. Taking all these considerations together leads to 
\bequa
  p_\mathrm{loose} &=&  \frac{1}{2} p(c_\mu\! =\!1; c_{\mu-1}\!=\!1) + (1-p(c_\mu\!=\!1; c_{\mu-1}\!=\!1) \nonumber \\
  &\approx& 0.39 .
\eequa
Here, we used the definition of the conditional probability, $p(c_\mu\! =\!1; c_{\mu-1}\!=\!1)=p(c_\mu\!=\!1|c_{\mu-1}\!=\!1)\times p(c_\mu\!=\!1)$.

Knowing the probability of a site being occupied by a loose particle, it seems tempting to subtract the corresponding occupations from the filling density $\rho_\mathrm{filling}$. But particle rearrangements in a dynamic system will  continuously create new loose particles. Hence, we have to replace sites occupied by a loose particle with sites of a suitable effective density $\rho_\mathrm{eff}$.  

To this end, consider a hole (i.e. an empty site) that propagates from right to left through the system. In each hopping step, the hole might either hit a column that contains a loose particle with probability $p_\mathrm{loose}$ or one that contains only stable particles with probability $1-p_\mathrm{loose}$. As only loose particles contribute to a modification of the density we are only interested in the first case.
The time between two consecutive holes arriving at any given position in the steady state will on average be $1/\beta$.  Therefore, a fully occupied column that contains a loose particle will on average remain in this state for a time $1/\beta$ before a hole arrives. Then, the hole either hits the loose or the stable particle in this column. If the stable particle is removed, the density is not changed and the column remains at full occupation until the next hole arrives, i.e. for another time $1/\beta$. Opposed to that, the column remains at half occupation for a time $1/\beta$ if the loose particle is removed. This leads to an effective (time-averaged) density of
\bequa
 \rho_\mathrm{effective} = \frac{\beta}{2}\cdot (\frac{1}{\beta} + \frac{1}{\beta} \cdot (\frac{1}{2}\cdot \frac{1}{2} + \frac{1}{2} \cdot 1 ) = \frac{7}{8}.
\eequa
Hence, the filling density $\rho_\mathrm{filling}$ and the maximal dynamic system density $\rho_\mathrm{max}$ are related via
\bequa
 \rho_\mathrm{max}\approx \rho_\mathrm{filling} - p_\mathrm{loose} ( 1 - \rho_\mathrm{effective}) .
\eequa
The result is $\rho_\mathrm{max}\approx 0.74$ which is in very good agreement with stochastic simulations that yield  $\rho_\mathrm{max}\approx 0.73$. This validates our understanding of topological hindrance for jammed configurations. 

Due to  approximations in the above calculation our computed value for $\rho_\mathrm{max}$ is, opposed to the one for $\rho_\mathrm{filling}$, not an exact result. Specifically, we implicitly assume that for non-vanishing $\beta\rightarrow 0$ occupations are created in the same way as they are in the filling process $\beta= 0$. The good quality of our result therefore indicates that, as a hole propagates through a jammed system, empty lattice sites are filled by particles in a mostly uncorrelated fashion. Note that the filling density does, in principle, depend on the particle in rate $\alpha$. If particles are injected with a high frequency mutual interactions while traveling to the jammed end lead to correlations and sorting effects that ultimately cause varying values for $\rho_\mathrm{filling}$. This dependency of the filling density on $\alpha$ is, however, irrelevant for our computation of the maximal (dynamic) density $\rho_\mathrm{max}$: Correlations that build up during the particles' motion can also be neglected in the case $\beta \rightarrow 0$ as the system is almost completely jammed.

\begin{figure}[t]
\centering
\includegraphics[width = \columnwidth]{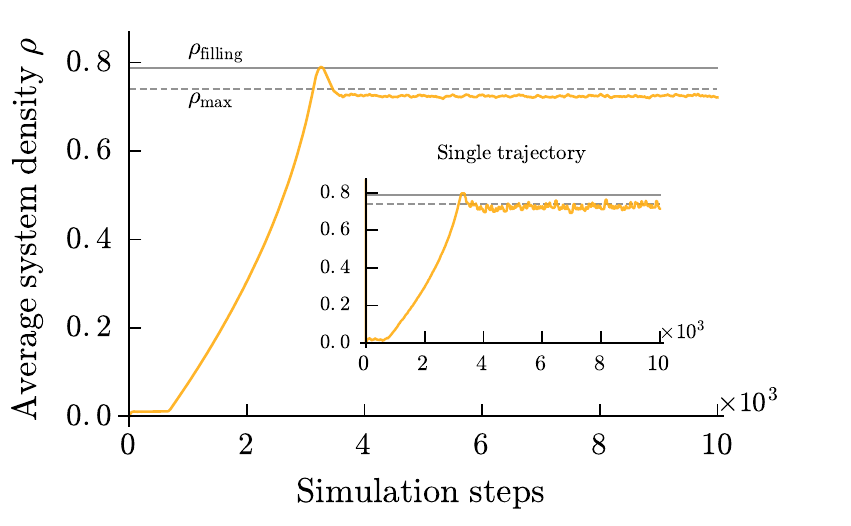}
\caption{\textbf{Temporal evolution of the system density for $\alpha\rightarrow 0, \beta\rightarrow 0, \beta \ll \alpha$.} As the number of simulation steps increases (x-axis), particles enter the system, ``pile up'' at the end and create an average density close to $\rho_\mathrm{filling}\approx 0.79$ (solid line). As soon as particles leave the system the presence of loose particles is suppressed which leads to a drop of the average density until the maximal dynamic density $\rho_\mathrm{max}\approx 0.73$ is reached. The dashed line depicts the result of the analytic approach for $\rho_\mathrm{max}$ described in this section. Here, the density is obtained from an ensemble and spatial average. A single realization (inset) shows that the average system density (i.e. spatial average) oscillates around $\rho_\mathrm{max}$, which can be explained by large reordering events if stable particles are removed after several loose particles were extracted. Simulation parameters: $\alpha= 0.02$, $\beta=10^{-4}$, $\delta=1/2$, $W=2$, $L=4096$.}
\label{fig:DensityOvershoot}
\end{figure}

The theoretical arguments of this section addressing a transition from the filling density to the maximal dynamic density are further illustrated with the following consideration. Fig.~\ref{fig:DensityOvershoot} shows the temporal evolution of the average system density with a very small out rate, starting from an empty system. The out rate was chosen such that, on average, the system completely fills with particles before a first particle exits. Indeed, we observe that the density gradually increases until $\rho_\mathrm{filling}$ is reached. In particular, the density overshoots the maximal dynamic density $\rho_\mathrm{max}$. As soon as particles leave the system, loose particles are removed which causes the average density to decrease and fluctuate around $\rho_\mathrm{max}$. These fluctuations are also explained by our above argumentation: As single holes travel through the system particles are reordered and loose particles are transiently created which leads to an (also transient) increase of the system density. 

\section{Derivatives of the hindrance function  \label{ap:derivatives_of_hindracne}}
In the following, we are going to calculate the derivatives of the hindrance function for the two extremal cases $\rho \rightarrow 0$ and $\rho \rightarrow \rho_\text{max}$, as used in Section~\ref{sec:current_density}. To do so, we determine the change in current d$J$ in response to a change in density d$\rho$ at a given lattice site and equate it with the derivative of the current-density relation, Eq.~\ref{eq:current_density},
\begin{align}
\dfrac{\text{d}}{\text{d}\rho} J(\rho)=1- H(\rho)-\rho H'(\rho)-2\rho.\label{eq: appendix_current_derivative}
\end{align}
In the mean-field approximation a lattice site is occupied with a particle with probability $\rho$ irrespective of the neighboring lattice sites. We find an $S$ particle at any given lattice site with probability $\delta\rho$ and a $T$ particle with probability $(1-\delta)\rho$, respectively. 
Increasing the density by a value d$\rho$ we have two contributions to the change of the current. First, a decrease since particles are more like to be blocked, and, second, an increase because the additional density itself may contribute to the overall current. Both contributions can again be split into two separate cases. Either we are dealing with a focused state, see Fig.~\ref{fig:hindrance_illustration} (b) group 1, where two particles are blocked simultaneously or the case where one particle is affected only. Two particles from neighboring lanes  are focused on the same lattice site with probability $\rho^2  \delta (1-\delta )$ resulting in a current reduction of $-\rho^2  \delta (1-\delta )$d$\rho$. A single particle is blocked with probability $\rho \delta (1-\rho (1-\delta ))  +(1-\delta )\rho(1-\delta\rho)$. Here, $\rho \delta (1-\rho (1-\delta ))$ is the probability of only one $S$ particle being blocked and $(1-\delta )\rho(1-\delta\rho)$ only one $T$ particle being blocked. The same contributions can be derived for the increase in current due to the removal of a single particle. Summing all up we obtain 
\begin{align}
\label{eq:current_change_mf}
\dfrac{\text{d}}{\text{d}\rho} J =& - \rho^2  \delta (1-\delta )-\rho \delta (1-\rho (1-\delta )) \\ & -(1-\delta )\rho(1-\delta\rho)+(1-\delta )\delta\rho (1-\rho)\nonumber  \\ &+(1-\delta )(1-\delta\rho)(1-\rho)+\delta (1-(1-\delta )\rho)(1-\rho)\nonumber  \\
=&(1-2\rho)(1-\delta(1-\delta)\rho ).\nonumber
\end{align}
This expression reduces to the derivative of the TASEP current-density relation for $\delta=0$ and $\delta=1$.

The mean-field result is not correct in general for our system but becomes exact in the limit $\rho \rightarrow 0$.
Using Eq.~\ref{eq: appendix_current_derivative} our result for the derivative of the hindrance function reads
\begin{align}
1- H(\rho)-\rho H'(\rho)-2\rho \overset{!}{=}  (1-2\rho)(1-\delta(1-\delta)\rho ).
\end{align}
As, by definition, $H(\rho)/ \rho \rightarrow H'(0)$ for $\rho \rightarrow 0$, we obtain 
\begin{align}
H'(0)&=\delta(1-\delta).
\end{align}

We proceed in a similar way for the case $\rho \rightarrow \rho_\text{max}$. However, instead of an uncorrelated system we face total correlations between particle occupations; A site is only occupied if the one in front is as well. Filling the last hole in a jammed configuration will not increase the current but prevent the last possible particle motion. This means that we are left with those parts of Eq.~\ref{eq:current_change_mf} that have a negative contribution. Evaluating the corresponding equation at the maximal density $\rho_\mathrm{max}$ we find
\begin{align}
\dfrac{\text{d}}{\text{d}\rho} J =&-\delta\rho_\mathrm{max} (1-\rho_\mathrm{max} (1-\delta))\nonumber \\ &-(1-\delta)\rho_\mathrm{max} (1-\delta \rho_\mathrm{max})- \delta(1-\delta) \rho_\mathrm{max}^2 \\
=& -\rho_\mathrm{max} +\delta (1-\delta) \rho_\mathrm{max}^2.\nonumber
\end{align}
Again, using Eq.~\ref{eq: appendix_current_derivative} and the definition of the maximal density $H(\rho_\mathrm{max})=1-\rho_\mathrm{max}$ we obtain
\begin{align}
 H'(\rho_\mathrm{max})&= -\delta(1-\delta) \rho_\mathrm{max}, \nonumber
\end{align}
for the derivative of $H(\rho_\mathrm{max})$. Note that, opposed to the limit $\rho\rightarrow 0$, this is not an exact result but should be interpreted as a refined mean-field approximation. 

\section{Derivation of current and density response to the control parameters in the high- and low-density phases\label{ap:phase_diagram}}
In this section we derive the dependence of the current and density on the in rate $\alpha$ and out rate $\beta$. As for the TASEP, collective behavior in the low-density phase only depends on the in rate $\alpha$ whereas the high-density phase is determined by the out rate $\beta$.
\subsection{The low-density phase}
\begin{figure}[t]
\centering
\includegraphics[width = \columnwidth]{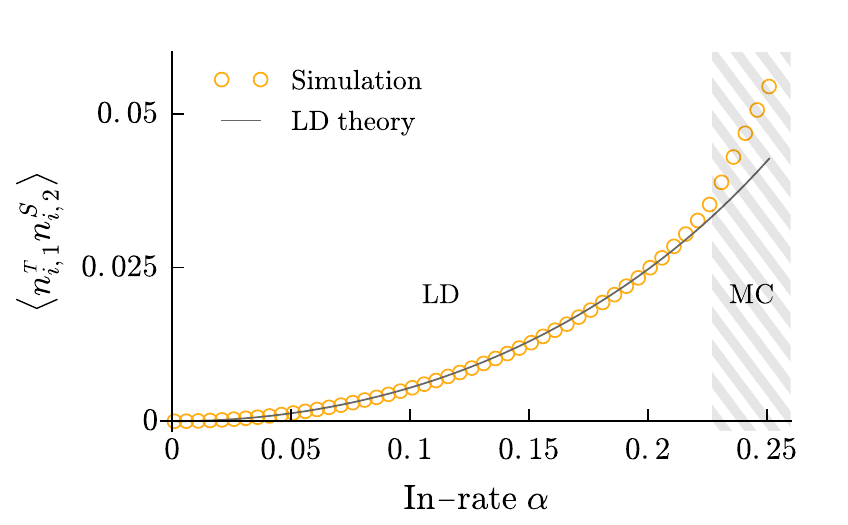}
\caption{\textbf{Second moment $\avg{n^T_{i,1} n^S_{i,2}}$ of occupations of different species at the first two lattice sites as predicted by our theory (lines) and stochastic simulations (symbols) for the special case $\delta=1/2$ and $W=2$.} Deviations between theory and simulation only occur close to the phase transition at $\alpha \approx 0.22$.
 \label{fig:LD_correlations}}
\end{figure}

For many driven lattice gases, current conservation is used to relate the system's current and density to in and out rates. For the case of the TASEP, the low-density phase is dictated by an influx $J=\alpha (1-\rho_1)$. Further, there is no boundary layer at the left lattice end in the low-density phase such that the density at the first site equals the bulk density. Thus, due to current conservation, $\rho_1 = \alpha$, for the TASEP. For our model, however, we face a new problem. Simulations suggest that, even in the low-density phase, the occupation of a first lattice site is not equal to the bulk density and is also not identical to the reservoir density $\alpha$. This is due to non-trivial correlations between occupations at the left lattice end, see Fig.~\ref{fig:LD_correlations}, which we will compute in the following.
Indeed, the density at the first site is always slightly higher than $\alpha$. Also, when, for example, the in rate of $T$ particles $\alpha_T$ is increased the density of  $S$ particles at a first lattice site $\rho^S_{i,1}$ increases and vice versa. The latter observation, in particular, contradicts equating $\rho^S_{i,1}=\alpha_S$.
This behavior can be understood in the following way. Consider, for example, a $T$ particle at the first lattice site. This $T$ particle can never prevent an $S$ particle to hop in front of it; The $T$ particle can be blocked by the $S$ particle but can not block the $S$ particle. Therefore, this $T$ particle will stay on average longer at the first lattice site as compared to the TASEP and the corresponding density $\rho_{i,1}$ will be higher than $\alpha$. The exact flux-balance condition for $T$ particles, Eq.~\ref{eq:master_in_T}, reads
 \begin{eqnarray}
 \label{eq:flux_balance_LD}
  \alpha_T (1-\avg{n^T_{i,1}} \!- \! \avg{n^S_{i,1}}) &=& \avg{n^T_{i,1}} \! -\!  \avg{n^T_{i,1} n^T_{i,2}} \! - \! \avg{n^T_{i,1} n^S_{i,2}} . \nonumber \\
 \end{eqnarray}
Again, correlators of the form $\avg{n^{X_1}_{i,1} n^{X_2}_{i,2}}$ lead to an unclosed set of equations. To overcome this problem, we approximate second moments. Note that all particles enter the system from the left reservoir in an uncorrelated fashion. The way $T$ particles can block and are blocked by other  $T$ particles follows the same scheme as in the TASEP. Therefore, it is reasonable to assume correlations to balance (as for the TASEP). Thus we factorize $\avg{n^T_{i,1} n^T_{i,2}}\approx \avg{n^T_{i,1}}\avg{n^T_{i,2}}$ for the same species. This is different for $\avg{n^T_{i,1} n^S_{i,2}}$, where a TASEP like balance of blockage end getting blocked no longer holds true as argued above.

Explicitly writing down the master equation for $\avg{n^T_{i,1} n^S_{i,2}}$ in the steady state leads to
 \begin{eqnarray}
  \frac{\mathrm{d}}{\mathrm{dt}} \avg{n^T_{i,1} n^S_{i,2}}\! &=&\! 0  = \alpha_T \avg{(1\! -\! n^T_{i,1} \!- \! n^S_{i,1}) n^S_{i,2} } \nonumber \\
  && + \avg{n^T_{i,1} n^S_{i-1,1} (1-n^T_{i,2} \! - \! n^S_{i,2})} \nonumber \\
  && - \avg{n^T_{i,1} n^S_{i,2} (1-n^T_{i+1,3} \! - \! n^S_{i+1,3})} .
 \end{eqnarray}
As particles that enter the system are uncorrelated, we can set  $\avg{n^T_{i,1} n^S_{j,1}} = \rho^T_{i,1} \rho^S_{j,1}$ for $i \neq j$. Further, we are interested in a solution that is valid in the low-density phase, where we expect $\alpha$ to be small and $\rho_{i,1}$ to be of order (but not necessarily equal to) $\alpha$, i.e. $\rho_{i,1} \propto \mathcal{O} (\alpha)$. If correlations are week, we can also assume that $\avg{n_{i,\mu} n_{j,\nu}}\propto \mathcal{O} (\alpha^2)$, and $\avg{n_{i,\mu} n_{j,\nu} n_{k,\xi}}\propto \mathcal{O} (\alpha^3)$. The transitions from the low-density phase to the maximal-current  and high-density phases take place at lower values of $\alpha$ as compared to the TASEP (cf. Fig.~\ref{fig:current_density_multilane}) which ensures very low values for the density in the low-density phase in our model.
 Then, upon truncating at third order in $\alpha$, we arrive at an equation for $\avg{n^T_{i,1} n^S_{i,2}} $ that is valid for low densities:
  \begin{eqnarray}
  \avg{n^T_{i,1} n^S_{i,2}} &\approx& \alpha_T \rho^S_{i,1} + \rho^T_{i,1} \rho^S_{i,1}.
 \end{eqnarray}
Here, we also assumed that $\rho^S_{i,2} \approx \rho^S_{i,1}$~\footnote{Note that the site index $\mu$ was kept here mainly to keep track of the correct form of correlations between sites and lanes rather than due to a rapid spatial change of correlations or densities.}. An according equation can be derived for $\avg{n^S_{i,1} n^T_{i,2}}$. Inserting the results in Eq.~\ref{eq:flux_balance_LD} and its analogue for the influx of $S$ particles we find 
 \begin{subequations}
 \label{eq:LD_entry_eqns}
\begin{eqnarray}
  \alpha_T (1-\rho_{i,1}^T) &=& \rho_{i,1}^T (1- \rho_{i,1}^T - \rho_{i,1}^S) \\ 
  \alpha_S (1-\rho_{i,1}^S) &=& \rho_{i,1}^S (1- \rho_{i,1}^T - \rho_{i,1}^S) .
\end{eqnarray}
\end{subequations}
Eqs.~\ref{eq:LD_entry_eqns} can be interpreted as a two-species TASEP where particles from different species can not block each other when entering the system. This means that, for low densities,  the possibility of overtaking particles from the other species when entering the system effectively acts as if no exclusion was present for different species;  Particles from the same species still exclude each other. Eqs.~\ref{eq:LD_entry_eqns} can be solved for$\rho^T_{i,1}$ and $\rho^S_{i,1}$ which fixes the current and bulk density in the system.

For the case of identical in rates the solution of Eqs.~\ref{eq:LD_entry_eqns} is of a particularly simple form and the total current is given by
\begin{equation}
J=\dfrac{\alpha}{4}\left( 2-\alpha+\sqrt{4+(\alpha -12)\alpha }\right) .
\end{equation}
\begin{figure}[t]
\centering
\includegraphics[width = \columnwidth]{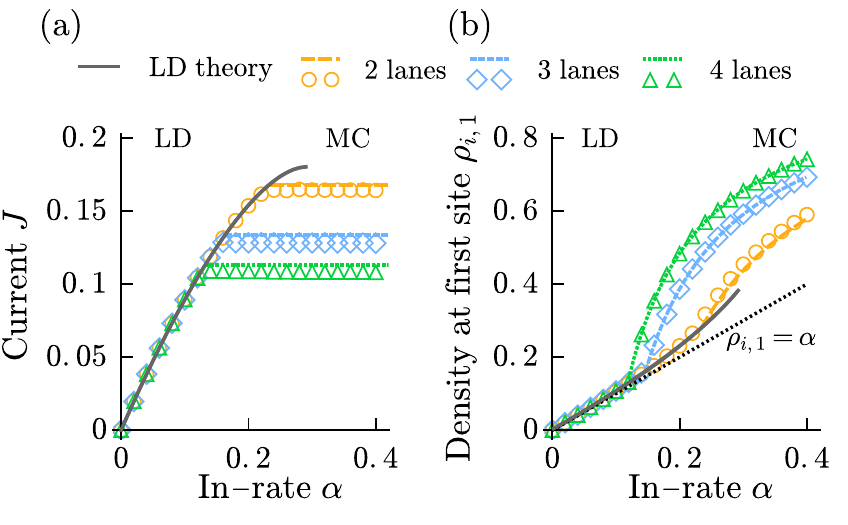}
\caption{\textbf{Comparison between low-density theory (lines) and stochastic simulations (symbols) for the special case $\delta=1/2$}\; \textbf{(a)} Current plotted against the in rate $\alpha$ in the low-density phase (LD) for different numbers of lanes $W$. For low in rates the current increases monotonically until it becomes constant at a point that depends on the number of lanes; The respective current corresponds to the maximal-current  of the bulk.  Low-density theory agrees with simulation results for the low density phase. The low-density theory is independent of the number of lanes. \textbf{(b)} Density at the first lattice site $\rho_{i,1}$ plotted against the in rate $\alpha$ for different numbers of lanes.
In the low-density phase the density is dictated by the left boundary and agrees with the low-density theory. After the transition to the maximal-current  phase the density is determined by conservation of the maximal current. Again, the corresponding value for the maximal current depends on the respective number of lanes which leads to a distinct behavior for different numbers of lanes in the MC phase.\label{fig:LD_current}}
\end{figure}

Even though this result is independent of the number of lanes the actual current in the system can never become higher than the respective maximal-current of the bulk that, in turn, depends on the number of lanes. As soon as the bulk limits transport the transition to the maximal-current  phase is triggered. For higher in rates the conservation of the maximal current dictates the density at the first lattice sites. A comparison with  simulation data for different numbers of lanes is shown in Fig.~\ref{fig:LD_current}. 
The  low-density theory agrees with the data up to the point of the phase transition. The exact value of the in rate $\alpha$ at which the transition into maximal current phase takes place depends on the number of lanes.

In summary, Eq.~\ref{eq:LD_entry_eqns} correctly describes behavior of our system in the low-density phase that is found to be independent of the number of lanes $W$ of the system. Dependencies on the number of lanes are only relevant for phase transitions which can be derived by applying the extremal current principle.

\subsection{The high-density phase \label{ap:high_density}}
\begin{figure}[t]
\centering
\includegraphics[width = \columnwidth]{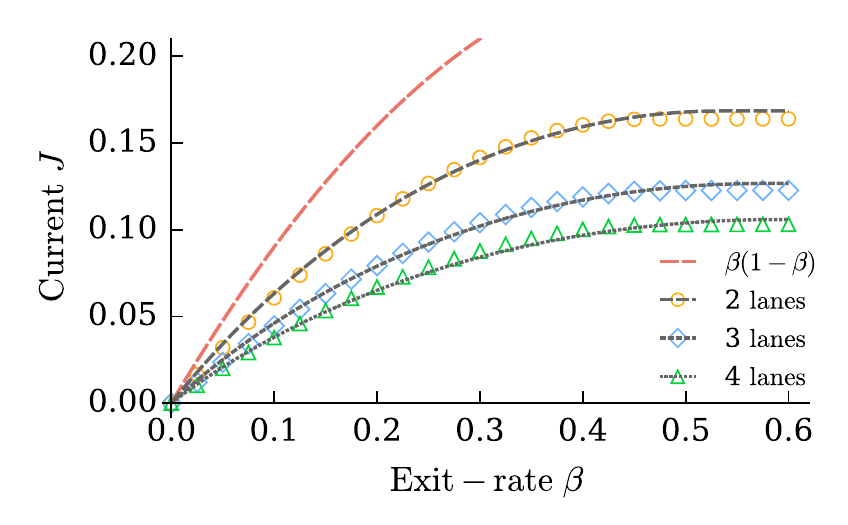}
\caption{\textbf{Average system current as a function of the out rate $\beta$ in the high-density phase for different numbers of lanes for the special case $\delta=1/2$.}  Comparison between stochastic simulations (symbols) and theoretical predictions (lines). The current increases monotonically with increasing out rate. For sufficiently high out rates the system enters the maximal-current  phase and the current becomes constant. The assumption of $\rho_L=1-\beta$ with a corresponding out flux of $J=\beta (1-\beta)$ (uppermost dashed line) fails to describe our model. \label{fig:HD_current}}
\end{figure}

In the high-density phase the system's current is dictated by the right boundary and determined by an outflux of particles $J=\beta \rho_{L}$. Here $\rho_{L}$ denotes the density at the last lattice site for an arbitrary lane $i$. Due to current conservation, this outflux must equal the bulk current,
\begin{align}
\beta\rho_L&= \rho(1-H\left(\rho\right)-\rho),\label{eq:HD_curr1}
\end{align}
where  $\rho$ denotes the density in the bulk. As for the low-density phase we lack knowledge about the value of the density at the dominating boundary, in this case $\rho_L$. Clearly, the mean-field result $\rho_L = 1-\beta$ can not be used. This becomes evident when we consider jammed systems that exhibit a  maximal density $\rho_\mathrm{max}$ that does, in general, not equal $1-\beta$. However, we expect the density at the right lattice end to be of the order of the bulk density, $\rho_L\approx \rho$. Using this assumption in Eq.~\ref{eq:HD_curr1} closes the equation. The result is shown in Fig.~\ref{fig:HD_current} and is in agreement with stochastic simulations.

\section{Supplemental simulations}
In the following section we present additional data from simulations that supplement several statements made in the main text.
\subsection{Rotational symmetry of the density profile \label{ap:Rotational_Invariance}}
\begin{figure}[h]
\centering
\includegraphics[width = \columnwidth]{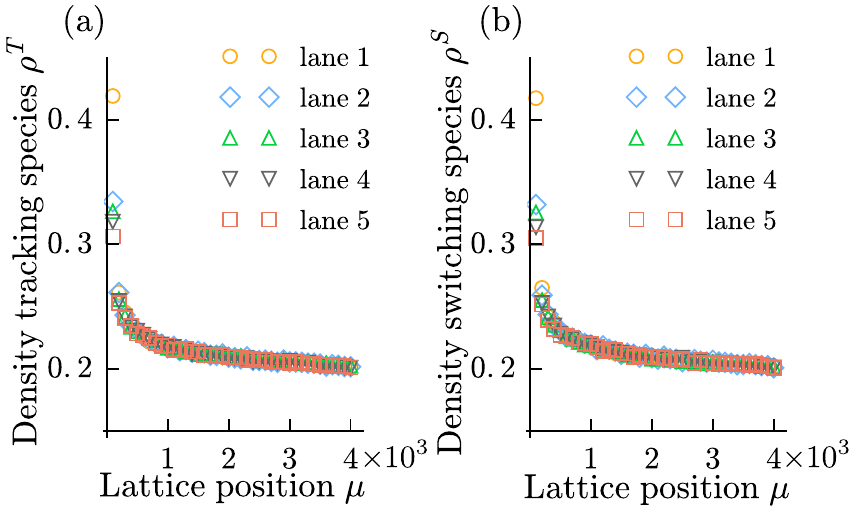}
\caption{\textbf{Comparison of the density profiles of all lanes for a system with $W=5$ lanes.} The density profiles are identical for each lane and for both species. No symmetry breaking takes place. In order to improve visibility only every 100-th lattice site is plotted for every lane with a different offset for each lane.  Other parameters: $L=4096$, $\delta=0.5$, $\alpha=10^{-4}$, $\beta=10^{-6}$.}
\label{fig:Rotational_Invariance_Density_profile}
\end{figure}

In the main text we discussed that density profiles have to be identical for every lane because any stationary state has to be unique. However, the time needed to reach this stationary state can be very long since the system may get stuck in meta-stable configurations that break symmetry.  Fig.~\ref{fig:Rotational_Invariance_Density_profile} shows that indeed all lanes admit identical density profiles in our simulations. This demonstrates that simulation times are chosen sufficiently long to ensure convergence to the (rotationally invariant) stationary state.

\subsection{The relation between the species fraction of the current and the density \label{ap:species_fraction}}

\begin{figure}[h]
\centering
\includegraphics[width = \columnwidth]{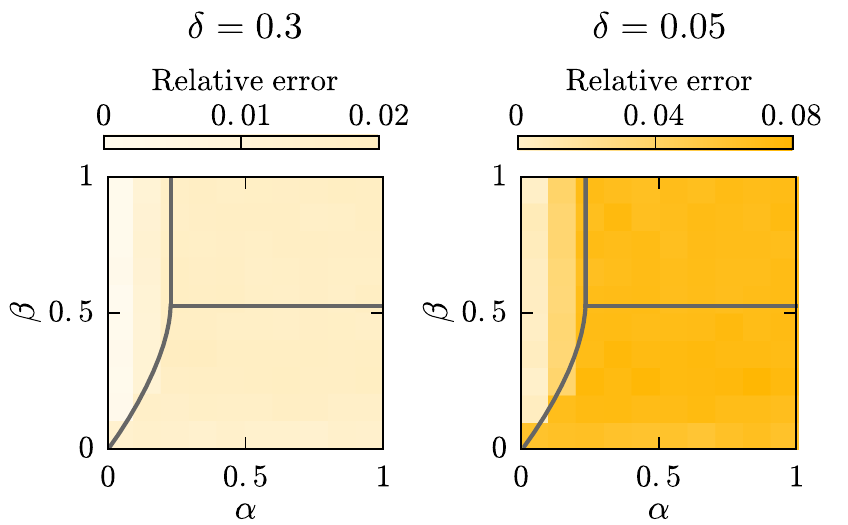}
\caption{\textbf{Relative deviation of the density species fraction from $\delta$ for the special case $W=2$ lanes and $L=4096$ sites.} Color denotes simulation results for the relative deviation and thus the error that is made when assuming a density species fraction of value $\delta$. Lines indicate the phase boundaries as predicted by our theory.  }
\label{fig:Error_Species_Fraction_2_lanes}
\end{figure}
\begin{figure}[h!]
\centering
\includegraphics[width = \columnwidth]{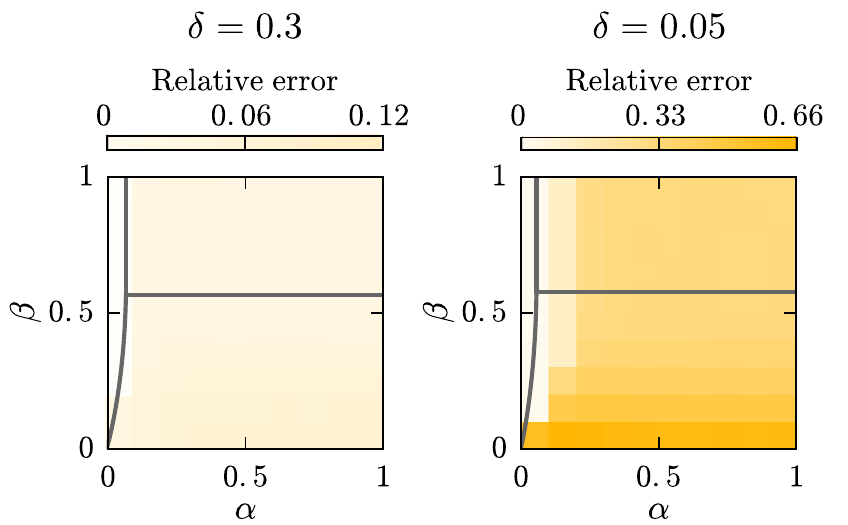}
\caption{\textbf{Relative deviation of the density species fraction from $\delta$ for the special case $W=13$ lanes and $L=512$ sites.}  Color denotes simulation results for the relative deviation and thus the error that is made when assuming a density species fraction of value $\delta$. Lines indicate the phase boundaries as predicted by our theory.}
\label{fig:Error_Species_Fraction_13_lanes}
\end{figure}

In section~\ref{sec:tophind}  of the main text we have shown that in the mean-field approximation the current species fraction $\delta$ is identical to the density species fraction $\rho^S / \rho$. This result does, however, not hold true in general. To asses the difference between the species fraction of the density and that of the current we have performed stochastic simulations for a wide range of parameter values. 
As illustrated in Fig.~\ref{fig:Error_Species_Fraction_2_lanes} our simulations show that deviations between both quantities are relatively small (below $10\%$). This justifies to use $\delta$ as the general species fraction in most cases and in our manuscript.

There are, however, important exceptions where this equivalence can not be employed. If we regard systems with a small aspect ratio $L/W$ and significantly asymmetric species mixtures the species fraction of current and density strongly differ. 
Fig.~\ref{fig:Error_Species_Fraction_13_lanes} shows differences of up to $60\%$ between these quantities in stochastic simulations with corresponding parameter values.
In all these cases the na\"{i}ve assumption $\rho^S\approx\delta\rho$ clearly fails and the individual densities have to be considered separately.

\vspace*{5cm}

\subsection{Current-density relation for a system with a large number of lanes\label{ap:current_density_13_lanes}}

\begin{figure}[h]
\centering
\includegraphics[width = \columnwidth]{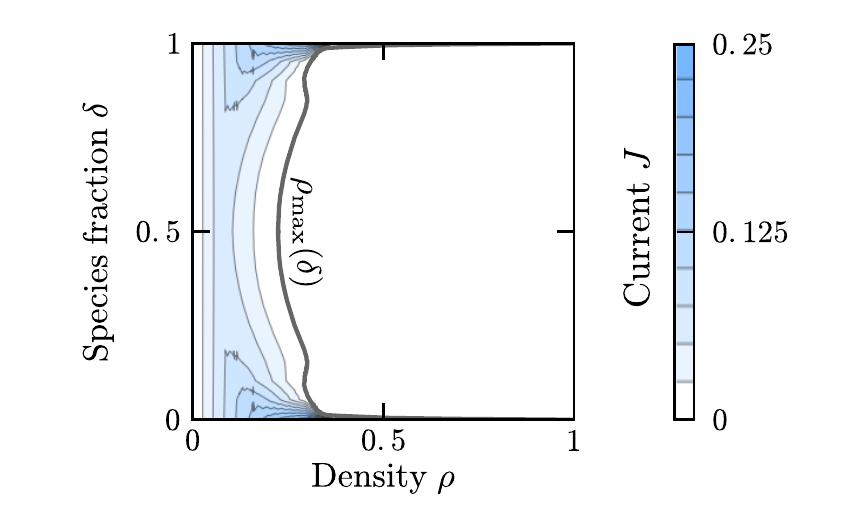}
\caption{\textbf{Current-density relation  obtained from stochastic simulations with $W=13$ lanes and $L=16384$ sites.}
The emerging current of an arbitrary lane (blue color) shows a dependence not only on the average system density but also on the fraction of lane-switching particles, $\delta$. The system jams at a maximal density $\rho_\mathrm{max}(\delta)$ (bold gray line); Densities between $\rho_\mathrm{max}$ and full occupation are not realized in the stationary state. }
\label{fig:current_denisty_13_lanes}
\end{figure}

The current density relation for a large number of lanes shows very drastic modifications as compared to the TASEP, see Fig.~\ref{fig:current_denisty_13_lanes}. Most of the possible density values are not realized at all in the stationary state. Further, the maximal current is far below the one known from the TASEP. As for low numbers of lanes already a very small fraction of the second species ($2-5\%$) is sufficient to drastically reduce transport efficiency. 

\subsection{Robustness of the density patterns with respect to model extensions\label{ap:robustness}}
\begin{figure}[h]
\centering
\includegraphics[width = \columnwidth]{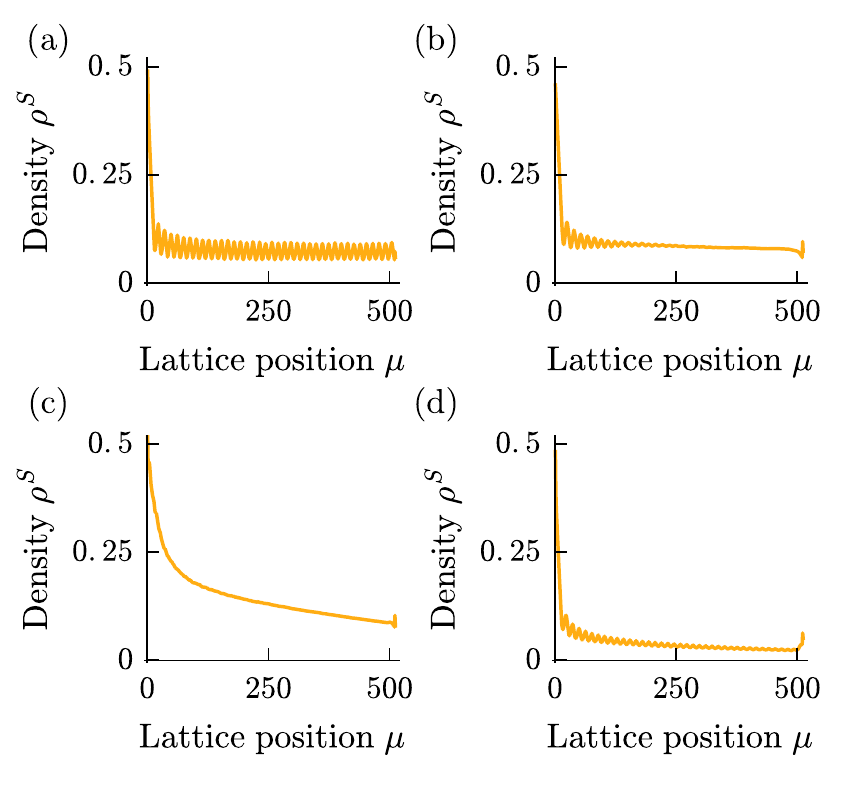}
\caption{\textbf{Effects of model extensions on density patterns.} \textbf{(a)} Dimers show the same kind of density patterns as monomers. \textbf{(b)} If lane-switching dimers do not follow a unique trajectory ($\theta=0.95$) density patterns occur but are subject to spatial decay. \textbf{(c)} As soon as dimers are allowed to attach randomly on the lattice ($\omega_a^S=\omega_a^T=5\cdot 10^{-5}$) the system loses its characteristic length scale and the density profile becomes smooth. \textbf{(d)} Random dimer detachment ($\omega_d^S=\omega_d^T= 10^{-3}$) still allows for the formation of density patterns. While the wavelength stays the same the amplitude of the oscillations is reduced. Other parameters: $W=13$, $L=512$, $\alpha =0.6$, $\beta=0.2$, $\delta =1/2$.}
\label{fig:dens_pattern_robustness}
\end{figure}

One of the most intriguing features of our main model is the formation of  regular density patterns. However, while this is clearly an interesting phenomenon to study in the context of non-equilibrium physics it is an open question whether such patterns may occur in a biological systems.  While a rigorous and complete analysis of pattern formation in this context it out of scope of this work, we would like to provide an outlook on potential changes due to different model extensions. Fig.~\ref{fig:dens_pattern_robustness} shows how the density patterns change as compared to Fig.~\ref{fig:pattern}~(a) of the main text if different modifications are included.

Modelling of particles as dimers  does not significantly affect the formation of patterns (see Fig.~\ref{fig:dens_pattern_robustness}~(a)); They again appear robustly and with the same wavelength as before. Since no differences between monomers and dimers are observed we restrict our discussion to the dimeric case for the rest of this section. 

 Including non-strict lane hopping leads to a spatial decay of oscillatory patterns (Fig.~\ref{fig:dens_pattern_robustness}~(b)). Opposed to that random particle attachment smoothens the density profiles globally (Fig.~\ref{fig:dens_pattern_robustness}~(c)). Finally, random particle detachment (even for comparably high detachment rates) still results in oscillatory patterns (Fig.~\ref{fig:dens_pattern_robustness}~(d)). This can be explained in the following way: To establish density patterns, particles of different species need to jam periodically due to particles they have encountered before. For the main model the typical length scale for this to happen is the width of the lattice. In the case of non-strict lane hopping, the period of circulating once around the system is stochastic which gradually destroys the ``phase" of our density-oscillations. If we add random particle
detachment to the system, length-scales are not changed: Some particles may leave the system which makes jams less likely but does not alter the length scale on which those jams appear. Random particle attachment critically changes this phenomenology: A certain lattice position does no longer correspond to a fixed run length of a specific particle. Therefore, the stationary density profile becomes translationally invariant which, in turn, has to suppress density patterns when averaged over large time windows.

%

\end{document}